\documentclass[aps,prresearch,onecolumn,groupedaddress,nofootinbib,floatfix]{revtex4-2}

\usepackage{amsmath,amssymb,bm}
\usepackage{graphicx}
\usepackage{xcolor}
\usepackage[hypertexnames=false]{hyperref}
\usepackage{algorithm}
\usepackage{algpseudocode}
\usepackage{booktabs}
\usepackage{braket}
\usepackage{siunitx}
\usepackage[capitalise]{cleveref}
\usepackage{placeins}   

\newcommand{\rlease}{\textsc{RLEASE}}
\newcommand{\casci}{\text{CASCI}}
\newcommand{\casscf}{\text{CASSCF}}
\newcommand{\nevpt}{\text{sc-NEVPT2}}
\newcommand{\hf}{\text{HF}}
\newcommand{\dmrg}{\text{DMRG}}

\newcommand{\tparam}{\tau}
\newcommand{\sone}{s_1}
\newcommand{\Ecasci}{E_{\casci}}
\newcommand{\Enevpt}{E_{\nevpt}}
\newcommand{\Edmrg}{E_{\dmrg}}

\begin{document}

\title{RLEASE: Reinforcement Learning Efficient Active Space Engine}

\author{Etinosa~Osaro}
\email{eosaro@psiquantum.com}
\author{Abhishek~Mitra}
\author{Andrew~J. Jenkins}
\author{Kelsey~A. Parker}
\author{Robert~H. Lavroff}
\author{Verena~A. Neufeld}
\author{Arpan~Kundu}
\author{Arvin~Kakekhani}
\email{akakekhani@psiquantum.com}
\author{Dario~Rocca}
\email{drocca@psiquantum.com}
\affiliation{PsiQuantum, 700 Hansen Way, Palo Alto, CA 94304}

\date{May 2026}

\begin{abstract}
Selecting the active space for multireference electronic-structure
calculations is a long-standing bottleneck that often requires
expert chemical intuition and costly trial-and-error.
We introduce \rlease{} (Reinforcement Learning Efficient Active
Space Engine), a low-cost method for automatic, geometry-dependent
active-space selection.
A neural network predicts per-orbital diagnostic scores
($\hat{s}_1$) from inexpensive Hartree--Fock orbital descriptors,
and a learned threshold partitions orbitals into active and inactive
sets.
The threshold policy is optimized with proximal policy optimization,
using the discrepancy between \nevpt{} energies computed with the
selected active space and \dmrg{} reference energies as the reward.
After training, the same RLEASE-selected active spaces can be used
with multireference perturbation theory or composite coupled-cluster
energy estimators.
Despite being trained on a small set of molecules and geometries,
\rlease{} transfers to chemically diverse test systems, producing
compact active spaces and competitive potential-energy surfaces
relative to established entropy-based selectors.
Because deployment requires only inexpensive orbital descriptors and
neural-network inference, \rlease{} enables high-throughput
multireference workflows without molecule-specific retraining or
target-system pilot \dmrg{} calculations.
\end{abstract}

\maketitle

\section{Introduction}
\label{sec:intro}

Many chemically important problems, including covalent bond breaking,
transition-metal chemistry, biradical intermediates, and electronically
excited states, require a wave function that cannot be described
qualitatively by a single electronic configuration.
In such cases, a multiconfigurational description is needed to capture
the near-degeneracies and strong correlation that define the underlying
electronic structure~\cite{Roos16_book,Shepard12_108,Bartlett12_182}.
Methods such as complete active space self-consistent field
(CASSCF)~\cite{CASSCF0,CASSCF1,CASSCF2}, complete active space
configuration interaction (CASCI), and density matrix renormalization
group (DMRG)~\cite{White1992,Chan2011} provide such
multiconfigurational reference wave functions, while post-reference methods such as multireference configuration interaction (MRCI)~\cite{Buenker1974,Buenker1978}, multireference coupled cluster (MRCC)~\cite{Bartlett12_182}, multireference perturbation
theory (MRPT2), including strongly contracted $N$-electron valence
state perturbation theory (\nevpt{})~\cite{Angeli2001,NEVPT2-1,NEVPT2-2}, recover additional
correlation and improve quantitative accuracy.
This general strategy follows the standard logic of active-space
methods: first obtain a qualitatively correct multiconfigurational
reference, then recover the remaining correlation outside the active
space with a more approximate but less expensive treatment~\cite{Gagliardi-review}.

In the longer term, fault-tolerant quantum algorithms are expected to
enable the treatment of larger active spaces than are practical
classically, with polynomial scaling for quantum simulation and the
potential for exponential speedups in eigenvalue estimation with
algorithms such as quantum phase estimation (QPE)~\cite{Lloyd1996}.
Active-space methods are important for quantum
chemistry on quantum computers as well, since the qubit and gate requirements
are determined by the size of the orbital space being treated.
Restricting the problem to a chemically meaningful active space
therefore provides a practical way to focus quantum resources on the
strongly correlated part of the system.
For early fault-tolerant quantum computers, active-space, embedding, and fragmentation strategies are therefore expected to remain essential for reducing the quantum problem to a tractable correlated subspace~\cite{embedding-gagliardi}.

The choice of active space remains one of the main
difficulties in the application of multiconfigurational methods~\cite{Roos16_book,Roos11_3329}.
The complete active space ansatz is useful because, in many chemical
systems, the dominant strong correlation is confined to a chemically
meaningful subset of orbitals, so that only a reduced portion of the
full Hilbert space must be treated at the multiconfigurational level.
This simplification is effective only if the selected orbitals span the
relevant strongly correlated subspace.
If important orbitals are omitted, the resulting zeroth-order wave
function is qualitatively incomplete, and subsequent post-CAS treatments
may not recover the missing physics.
If too many orbitals are included, the cost of the active-space solver
rises rapidly, even when approximate solvers such as DMRG are used.
As recent reviews have emphasized, active-space construction therefore
remains a central obstacle to making multiconfigurational methods more
routine and broadly accessible~\cite{Gagliardi-review}. 
In practice, active-space construction relies on expert chemical
intuition and iterative refinement, an inherently subjective,
non-transferable process that breaks down entirely in high-throughput
or geometry-scanning workflows.

Automated approaches have been proposed, but each has a critical gap.
Entropy-based methods~\cite{Legeza2003,Rissler2006,Stein2016,Sayfutyarova2017}
extract single-orbital entropies from a pilot \dmrg{} calculation
and threshold them to select active orbitals, but the pilot
calculation is itself expensive and the threshold is chosen
empirically, with no guarantee that the selected space minimizes
downstream energy error.
Natural-orbital occupation numbers from MP2 or
coupled cluster degrade in strongly correlated regimes.
Machine-learning approaches~\cite{Jeong2020,golub2021machine} can predict a
\emph{fixed} active space per molecule or give orbital diagnostics to select an active space, but these approaches are decoupled from the actual
energy objective and fixed-space approaches are unable to adapt as geometry changes along a
potential energy surface.
More generally, these approaches share a common limitation:
active-space selection is treated as a preprocessing step,
\emph{decoupled from the downstream electronic-structure method},
so the selected orbitals are never directly optimized for
energy accuracy.

We take a different approach.
\rlease{}
(\textbf{R}einforcement \textbf{L}earning \textbf{E}fficient
\textbf{A}ctive \textbf{S}pace \textbf{E}ngine) frames
active-space selection as a \emph{learned, energy-driven
optimization problem with a tractable scalar policy}.
A neural network maps cheap Hartree--Fock (HF) orbital descriptors
to per-orbital importance scores $\hat{s}_1$; a single threshold
$\tparam$ (the policy) partitions orbitals into active and inactive
sets.
The threshold is optimized end-to-end via proximal policy
optimization (PPO), with the discrepancy between the \nevpt{} energy
on the predicted active space and a \dmrg{} reference
as the reward.
Because the reward is an actual energy difference, \rlease{}
directly optimizes what prior methods only approximate: the
accuracy of the energy of the downstream correlated calculation.
Because the policy is a single scalar threshold, the reinforcement-learning
problem remains low-dimensional, which is advantageous when only a
small training set is available.

Concretely, \rlease{}:
\begin{enumerate}
  \item predicts per-orbital $\hat{s}_1$ scores from HF-based descriptors,
        eliminating the need for a pilot \dmrg{} calculation;
  \item uses PPO-based reinforcement learning to optimize a molecule- and geometry-dependent        selection threshold against DMRG reference energies, coupling orbital selection             directly to downstream energy accuracy;
  \item generalizes from three training molecules, sampled across multiple geometries in            a minimal STO-3G basis, to a chemically
        diverse test set spanning main-group diatomics,
        polyatomics, open-shell radicals, and 3$d$ transition-metal
        hydrides;
    \item provides a single learned active space that serves three
            downstream methods: \nevpt{} and two additive-subtractive
            formalism (ASF) composite coupled-cluster methods,
            ASF-CCSD and ASF-CCSD(T); training uses only sc-NEVPT2
            energies as the reward signal, while ASF is attractive in
            fault-tolerant quantum computing settings because the
            expensive high-accuracy solver only needs to be applied to compute the
            active space energy. In contrast, perturbative post-CAS methods such as
            \nevpt{} generally require higher-order reduced density
            matrices and are therefore substantially more expensive; and
  \item deploys to new geometries with negligible overhead beyond a single Hartree–Fock             calculation, enabling high-throughput MR workflows without retraining.
\end{enumerate}
Specifically, ASF is a composite energy
estimator that combines a lower-scaling baseline treatment of the full
system with a higher-accuracy treatment of a selected active space,
while subtracting the corresponding baseline active-space contribution
to reduce double counting of correlation energy.
Within the ASF formalism, the energy of the full system is approximated as
\begin{equation}
E_{\mathrm{ASF}} = E^{L}_{\mathrm{full}}
+ \bigl(E^{H}(\mathcal{A}) - E^{L}(\mathcal{A})\bigr),
\end{equation}
where $L$ denotes a lower-scaling baseline method, $H$ a
higher-accuracy active-space solver, and $\mathcal{A}$ the
active orbital set.
In this way, ASF restricts the expensive high-accuracy calculation to
the strongly correlated active space while recovering the remaining
correlation from a cheaper full-system method.
This is particularly attractive in fault-tolerant quantum computing
settings, where the active-space contribution may be obtained from a
quantum solver such as quantum phase estimation, while the remaining
dynamic correlation is treated classically.
By contrast, post-active-space methods such as \nevpt{} generally
require higher-order reduced density matrices, which substantially
increase both classical and quantum resource requirements.

An overview of the \rlease{} framework is shown in
\cref{fig:schematic}.
The remainder of the paper is structured as follows:
\Cref{sec:theory} provides some general background on active space approaches and methods to recover dynamical correlation;
\Cref{sec:method} describes the \rlease{} framework;
\Cref{sec:comp} provides computational details;
\Cref{sec:results} presents and discusses numerical results;
\Cref{sec:conclusions} summarizes our conclusions.

\section{Theoretical Background}
\label{sec:theory}

\subsection{Active-space electronic structure methods}
\label{sec:theory:cas}

The complete active space (CAS) framework partitions the molecular
orbital space into inactive, active, and virtual
orbitals, and performs a full configuration interaction (FCI) expansion
within the active space of CAS($N_e$,$N_o$), where $N_e$ electrons occupy
$N_o$ space orbitals~\cite{CASSCF0,CASSCF1,CASSCF2}. 
This exactly captures static (multireference) correlation among the
active orbitals; dynamic correlation from the remaining electrons is
then recovered perturbatively (NEVPT2, CASPT2) or by MRCI.
For large active spaces where exact diagonalization is intractable,
DMRG replaces the FCI solver while maintaining controllable accuracy
through the bond dimension $D$~\cite{White1992,Chan2011}.
The accuracy of any CAS calculation is therefore governed by
the choice of active orbitals: a set that is too small produces a
qualitatively wrong zeroth-order reference that no post-CAS treatment
can repair, while one that is too large rapidly escalates computational
cost.
Identifying the minimal sufficient active space systematically and
automatically is the problem that \rlease{} addresses.

\subsection{Single-orbital entropy for active-space selection}
\label{sec:theory:s1}

Because RLEASE first learns to predict orbital importance before optimizing the active-space threshold, we briefly review the single-orbital entropy, a standard DMRG-derived diagnostic used to quantify orbital correlation and guide active-space selection.
The single-orbital entropy for orbital $i$ is defined as
\begin{equation}
  \sone(i) = -\sum_{\alpha=1}^{4} \omega_\alpha^{(i)}
              \ln \omega_\alpha^{(i)},
  \label{eq:s1_def}
\end{equation}
where $\omega_\alpha^{(i)}$ are the eigenvalues of the one-orbital
reduced density matrix.

The one-orbital reduced density matrix is obtained by tracing the full
$N$-electron density matrix over all orbitals except $i$.  Because
each spatial orbital admits four occupation states (empty, spin-up,
spin-down, doubly occupied), $\omega_\alpha^{(i)}$ sums to unity
and $\sone(i) \in [0, \ln 4]$.
An orbital with $\sone \approx 0$ is well described by a single
occupation number (doubly occupied or empty), whereas a large $\sone$
signals strong entanglement with the remaining orbitals, i.e.,
pronounced multireference character.

Legeza and S\'olyom~\cite{Legeza2003} introduced the use of
single-orbital and two-orbital entropies in DMRG calculations,
and Rissler et al.~\cite{Rissler2006} demonstrated that these
quantities provide a quantitative measure of orbital correlations
in molecular systems.
Stein and Reiher~\cite{Stein2016} subsequently built an automated
active-space selection protocol that thresholds $\sone$ values
obtained from a low-cost pilot DMRG calculation.

The principal drawback of entropy-based selection is that a pilot
DMRG calculation in a large orbital space is itself expensive.
RLEASE removes this bottleneck by training a neural network to
predict $\hat{s}_1$ from inexpensive Hartree--Fock descriptors,
thereby eliminating the need for a pilot DMRG at inference time.

\subsection{Recovering dynamic correlation: sc-NEVPT2 and additive-subtractive coupled cluster}
\label{sec:theory:dynamiccorr}

After active-space selection, the CAS wave function provides a multiconfigurational description of the strongly correlated orbitals, but the total energy must also include dynamic correlation from inactive occupied and external virtual orbitals. This point is central to RLEASE training: each trial active space generated by the reinforcement-learning policy is evaluated by constructing a CASCI reference and then adding the sc-NEVPT2 dynamic-correlation correction, yielding the energy used in the DMRG-referenced reward. At deployment, the same RLEASE-selected active space is used in two complementary post-CAS frameworks: sc-NEVPT2, which provides the multireference perturbative energy used during training, and an additive-subtractive coupled-cluster formalism, which replaces the coupled-cluster description of the active orbitals with CASCI while retaining full-space CCSD or CCSD(T) correlation.

The strongly contracted $N$-electron valence state perturbation theory
(sc-NEVPT2) is a second-order perturbative correction applied on top of
a \casci{} or \casscf{} reference wave function~\cite{Angeli2001,Angeli2002}.
The CAS treatment captures the dominant static (multireference)
correlation within the active space, while sc-NEVPT2 accounts for the
remaining dynamic correlation through excitations involving inactive
(core) and external (virtual) orbitals.

The total NEVPT2 energy is
\begin{equation}
  \Enevpt = \Ecasci + E^{\text{corr}}_{\text{NEVPT2}},
  \label{eq:nevpt2}
\end{equation}
where $\Ecasci$ is the CASCI energy and
$E^{\text{corr}}_{\text{NEVPT2}}$ is the second-order correction.

In the strongly contracted formulation, each excitation subspace is
represented by a single contracted perturber function constructed from
the CAS reference, leading to a compact and size-consistent
approximation that is free of intruder-state instabilities.
sc-NEVPT2 energies are evaluated using PySCF~\cite{PySCF2018}.


As a second post-CAS strategy, we define an additive-subtractive formalism (ASF) that incorporates multiconfigurational active-space effects into an otherwise single-reference coupled-cluster energy. While the specific ASF expression used here is introduced in this work, it follows the general logic of additive-subtractive composite correlation schemes used in related coupled-cluster contexts: the coupled-cluster contribution associated with a chosen orbital subspace is removed and replaced by a higher-level treatment of that same subspace. In our implementation, the RLEASE-selected active orbitals are treated by CASCI, while dynamic correlation outside the active space is retained through a full-space CCSD or CCSD(T) calculation.
Explicitly, the coupled-cluster-based ASF energy is
\begin{equation}
  E_\text{ASF} = E_\text{CC}^\text{full}
                 + \bigl(\Ecasci(\mathcal{A})
                        - E_\text{CC}(\mathcal{A})\bigr).
  \label{eq:asf}
\end{equation}
The three terms in this equation are evaluated in the same Hartree--Fock molecular-orbital basis and are defined as follows:
\begin{itemize}
  \item A \textbf{full-space} coupled-cluster energy
        $E_\text{CC}^\text{full}$ (CCSD~\cite{Bartlett2007} or
        CCSD(T)~\cite{Raghavachari1989}) correlating all
        electrons in all orbitals above the frozen core.
  \item A \textbf{CASCI} energy $\Ecasci(\mathcal{A})$ obtained by
        performing full configuration interaction within the
        selected active space $\mathcal{A}$, with all orbitals
        outside $\mathcal{A}$ held at their Hartree--Fock occupations.
  \item An \textbf{active-space-restricted} coupled-cluster energy
        $E_\text{CC}(\mathcal{A})$, in which only the
        $N_e$ active electrons occupying the $N_o$ active orbitals
        are correlated at the CC level; all other electrons are
        frozen at their Hartree--Fock occupations.  This calculation
        uses the same orbital basis and active-space definition as the
        CASCI in step~2, ensuring consistent subtraction of the
        single-reference contribution from the active orbitals
        in the ASF correction.
\end{itemize}

The term in parentheses in Eq.~\eqref{eq:asf} replaces the coupled-cluster description of
the active orbitals with the CASCI result: $E_\text{CC}$
is subtracted to remove the CC treatment of those orbitals, and
$\Ecasci$ is added to supply the multireference treatment instead.
When the active space captures the dominant static correlation,
$\Ecasci$ is more accurate than $E_\text{CC}$ for those
orbitals, while the full-space CC energy provides dynamic
correlation from the remaining (inactive) electrons.

We evaluate two variants: ASF-CCSD, where both $E_\text{CC}^\text{full}$
and $E_\text{CC}$ are computed at the CCSD level, and
ASF-CCSD(T), where both use CCSD(T).  In both cases the active space
is determined by RLEASE using the NEVPT2 reward signal; no
coupled-cluster calculations are required during training.
This makes ASF-CC a purely inference-time augmentation: RLEASE
selects the active space, and the ASF energy leverages
that selection within a coupled-cluster framework.

\begin{figure}[ht]
  \centering
  \includegraphics[width=\textwidth]{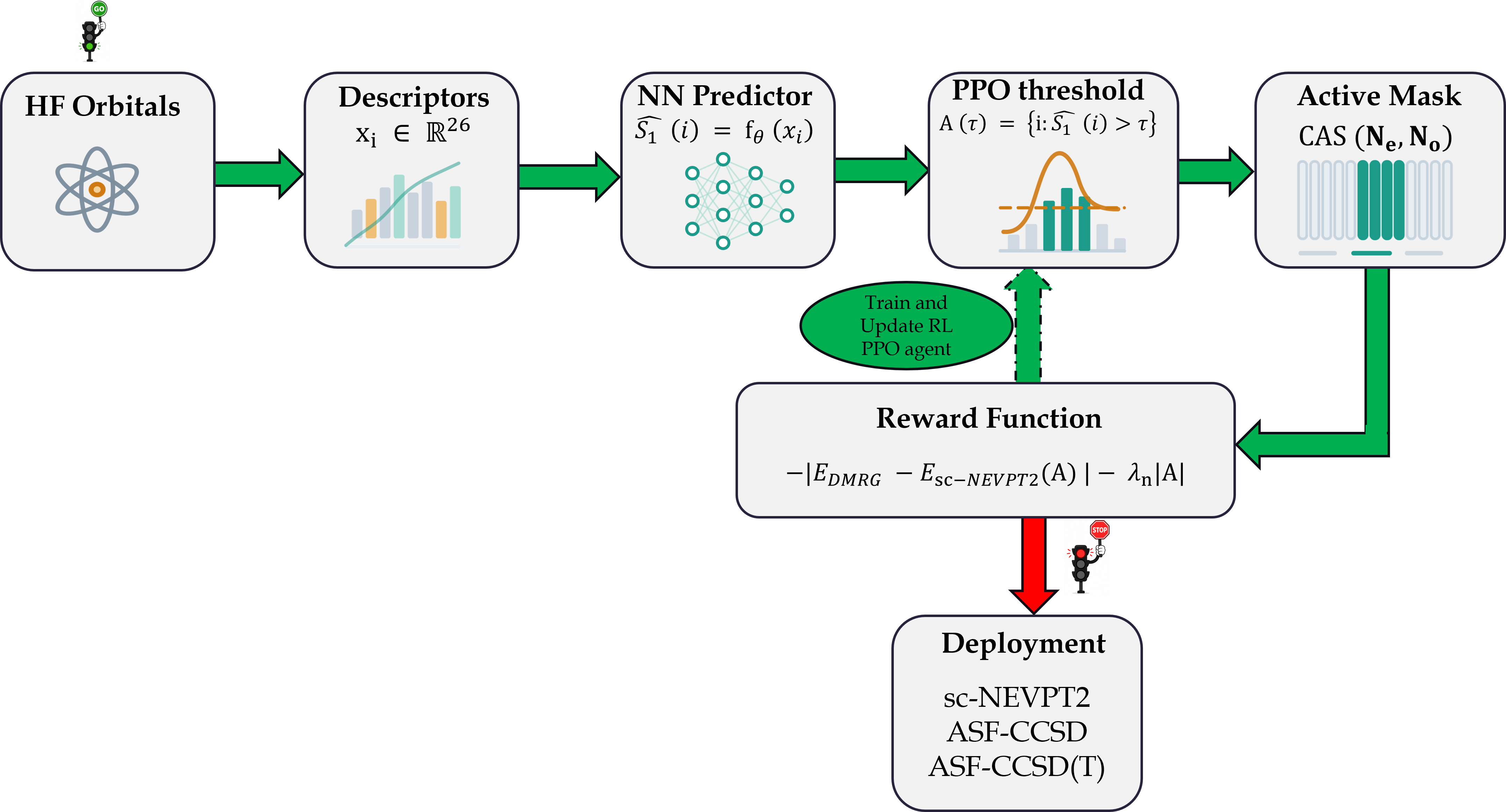}
  \caption{%
    Overview of the \rlease{} framework.
    Hartree--Fock orbital descriptors
    $\mathbf{x}_i \in \mathbb{R}^{26}$ are fed to a neural network
    that predicts per-orbital importance scores $\hat{s}_1$.
    A PPO-learned threshold $\tparam$ partitions orbitals into
    active and inactive sets, yielding a CAS($N_e$,$N_o$)
    specification.
    During training (green loop), the selected active space is
    evaluated via \casci{} + \nevpt{}, and the discrepancy with the
    \dmrg{} reference energy drives the RL reward signal that
    updates the threshold policy.
    At deployment, the same learned active space is used for three
    downstream methods: \nevpt{}, ASF-CCSD, and ASF-CCSD(T).
  }
  \label{fig:schematic}
\end{figure}

\section{The RLEASE Framework}
\label{sec:method}

The overall RLEASE workflow, from Hartree–Fock orbital descriptors to neural-network 
$\hat{s}_1$ prediction, PPO threshold optimization, active-space construction, and downstream energy evaluation, is summarized in Fig. \ref{fig:schematic}. Below, we describe each component in detail.

\subsection{Orbital descriptor construction}
\label{sec:method:descriptors}

For each molecular orbital $i$ obtained from a Hartree--Fock
calculation, we construct a feature vector
$\mathbf{x}_i \in \mathbb{R}^{26}$ composed of five groups of
physically motivated quantities:

\paragraph{Energetic features (4 components).}
The canonical orbital energy $\epsilon_i$;
the diagonal one-electron integral $h_{ii} = \langle i | \hat{h} | i \rangle$;
the diagonal two-electron self-repulsion integral
$g_{iiii} = (ii|ii)$;
and the spatial extent
$\langle r^2 \rangle_i = \langle i | \hat{r}^2 | i \rangle$.
Together, these features encode information about the energy scale and spatial delocalization
of each orbital.

\paragraph{Dipole magnitude (1 component).}
The magnitude of the orbital dipole vector,
$|\boldsymbol{\mu}_i| = |\langle i | \hat{\mathbf{r}} | i \rangle|$,
which measures the spatial asymmetry of the orbital charge distribution.
Using the magnitude rather than the Cartesian components ensures that
the descriptor is invariant under rigid rotations of the molecule.

\paragraph{Occupation and bonding labels (2 components).}
An occupation label $n_i \in \{0, 1, 2\}$ indicating doubly occupied,
singly occupied, or virtual character in the HF determinant; and a
bonding character label.  The bonding label is determined by summing the
overlap-weighted cross-atom MO coefficients: for each atom pair $(A,B)$
within 6~\AA{}, the score
$S_{AB} = \sum_{\mu \in A}\sum_{\nu \in B} c_\mu\, S_{\mu\nu}\, c_\nu$
is accumulated.  If one atom contributes $>95\%$ of the MO weight,
the orbital is classified as nonbonding (0); otherwise, a positive
total score gives bonding ($+1$) and a negative total score gives
antibonding ($-1$).

\paragraph{Atomic orbital composition (15 components).}
A binary encoding indicating which AO shell types contribute to MO~$i$.
The 15 entries correspond to the shells
1$s$, 2$s$, 3$s$, 4$s$, 5$s$, 2$p$, 3$p$, 4$p$, 5$p$, 3$d$, 4$d$,
5$d$, 4$f$, 5$f$, 5$g$; entry $j$ is set to 1 if any AO primitive
of that shell type has coefficient $|c_\mu| > 10^{-8}$ in the MO
expansion, and 0 otherwise.
This provides the network with $s/p/d/f$ angular-momentum character
without requiring explicit Mulliken analysis. 
Because the encoding records shell-type presence rather than the
number of basis functions, it reduces sensitivity to basis-set size.
However, transfer to basis sets containing shell types absent from
training (e.g., polarization functions not present in STO-3G)
remains a stress test of the model's generalization.

\paragraph{Approximate pair coefficient (APC)-derived features (4 components).~\cite{King2021,King2022}}
We include four features derived from the
APC active-space ranking scheme.
APC estimates orbital entropies from Hartree--Fock calculations
by constructing approximate pair coefficients between doubly occupied and
virtual orbitals from diagonal Fock and exchange matrix elements in the
molecular-orbital basis.
Here we do not use APC as the selector itself; instead, we include
per-orbital APC scores as features for the neural network.
The first two features are the standard APC sum and normalized average
(with a hard partition into occupied and virtual orbitals).
The remaining two are ``soft'' variants that replace the hard partition
with continuous occupation weights
$w_{\mathrm{occ},p} = n_p/2$, $w_{\mathrm{vir},p} = 1 - w_{\mathrm{occ},p}$
and combine occupied-like and virtual-like pair-entropy contributions
as $S_p = w_{\mathrm{occ},p}\, S_p^{\mathrm{occ}} + w_{\mathrm{vir},p}\, S_p^{\mathrm{vir}}$,
avoiding discontinuities for partially occupied orbitals in
open-shell systems.

\medskip\noindent
Each molecular geometry is represented as a variable-length sequence
of these 26-dimensional orbital feature vectors, one per MO.
The descriptor construction requires only quantities available from a
single Hartree--Fock calculation (orbital coefficients, integrals,
and the overlap matrix) and adds negligible cost beyond the SCF itself.
Descriptors of this type draw on ideas from orbital-based ML
representations used for predicting correlation energies and active
spaces~\cite{Jeong2020}.

\subsection{Neural network for \texorpdfstring{$\hat{s}_1$}{s1} prediction}
\label{sec:method:nn}

The $\hat{s}_1$ predictor is a feedforward neural network
$f_\theta: \mathbb{R}^{26} \to \mathbb{R}$ mapping descriptor vectors
to orbital diagnostic scores:
\begin{equation}
  \hat{s}_1(i) = f_\theta(\mathbf{x}_i).
  \label{eq:s1_pred}
\end{equation}
The architecture consists of $L$ fully connected hidden blocks.  Each
block contains a linear layer, a ReLU nonlinearity, and layer
normalization (LN), with optional dropout:
\begin{equation}
  \mathbf{h}^{(\ell+1)}
    = \text{LN}\!\bigl[
        \text{ReLU}\!\bigl(W^{(\ell)}\mathbf{h}^{(\ell)}
        + \mathbf{b}^{(\ell)}\bigr)
      \bigr],
  \;\; \ell = 0, \dots, L{-}1.
\end{equation}
Training minimizes a Smooth-$L_1$ loss between predicted $\hat{s}_1$ and
DMRG-derived $\sone$ values on a training set, with cosine annealing
learning rate schedule and early stopping on a held-out validation
set.
The features are standardized (zero mean, unit variance) and the
target $s_1$ values are transformed before training using
$\log(1+x)$ for numerical stability.
At inference, predictions are inverse-transformed back to dimensionless
$\hat{s}_1$ values before thresholding, state-vector construction,
and comparison with entropy-threshold baselines.
In the calculations reported here,
the model uses hidden dimension $H=256$, AdamW optimization, Smooth-$L_1$
loss with $\beta=0.05$, and 2000 supervised training epochs.

\subsection{Threshold-based active-space selection}
\label{sec:method:tau}

Given predicted scores $\{\hat{s}_1(i)\}_{i=1}^{N_\text{MO}}$ and a
threshold $\tparam$, the active space is defined as
\begin{equation}
  \mathcal{A}(\tparam) = \bigl\{i : \hat{s}_1(i) > \tparam \bigr\}.
  \label{eq:active_set}
\end{equation}
Several constraints are enforced post-selection:
\begin{itemize}
  \item A minimum active-space size $|\mathcal{A}| \geq N_\text{min}$
        (default 2 orbitals).
  \item Exclusion of the ``all-orbital'' active space
        ($|\mathcal{A}| < N_\text{MO}$), as full CAS becomes
        infeasible.
\end{itemize}

\subsection{Reinforcement learning for \texorpdfstring{$\tparam$}{tau} optimization}
\label{sec:method:rl}

We cast active-space selection as a reinforcement learning problem
in which an agent interacts with a quantum-chemistry environment.
The RL formulation is defined by the following components:

\paragraph{State (observation).}
At each training step the agent observes a molecular state vector
$\mathbf{s}$ constructed from the predicted orbital entropies
$\{\hat{s}_1(i)\}$ and the Hartree--Fock orbital descriptors
$\{\mathbf{x}_i\}$.
Specifically, $\mathbf{s} \in \mathbb{R}^{86}$ concatenates:
(i)~seven summary statistics of the $\hat{s}_1$ distribution (mean,
standard deviation, max, min, median, and the fractions of orbitals
exceeding 0.1 and 0.2);
(ii)~pooled orbital-descriptor statistics (mean, standard deviation,
and max of each of the 26 descriptor features across all orbitals,
giving $3 \times 26 = 78$ values); and
(iii)~a normalized molecule size (number of MOs divided by 60).
This yields a compact fixed-length vector that encodes both the
identity and the electronic structure of the molecule regardless of
the number of orbitals, and varies continuously along the potential
energy surface.
\paragraph{Action.}
The agent selects a continuous action $a = \tparam > 0$
that determines which orbitals enter the active
space via $\mathcal{A}(\tparam) = \{i : \hat{s}_1(i) > \tparam\}$.
The action is drawn from a state-conditioned Gaussian policy
parameterized by a two-hidden-layer feedforward network
$\pi_\phi$ (with GELU activations and layer normalization)
that maps the state $\mathbf{s}$ to a per-molecule mean threshold:
\begin{equation}
  \mu(\mathbf{s}) = \text{softplus}\!\bigl(\pi_\phi(\mathbf{s})\bigr),
  \qquad
  \tilde{\tparam} \sim \mathcal{N}\!\bigl(\mu(\mathbf{s}),\, \sigma^2\bigr),
  \qquad
  \tparam = \max(\tilde{\tparam},\, 0),
  \label{eq:tau_sample}
\end{equation}
where $\phi$ denotes the network weights, softplus ensures
$\mu > 0$, and $\sigma = \exp(\log\sigma)$ is a
learnable standard deviation (optimized in log-space and clamped
to $[10^{-4},\, 0.08]$).
The clipping at zero guarantees a non-negative threshold; in
practice the small $\sigma$ relative to $\mu$ makes this
binding constraint rarely active.
Because $\mu$ is a function of the state $\mathbf{s}$, the policy
produces a different threshold distribution for every molecule and
geometry, adapting to each system's characteristic $\hat{s}_1$ profile.
At deployment, the deterministic action $\tparam = \mu(\mathbf{s})$
is used.

\paragraph{Environment.}
The environment is a black-box quantum-chemistry simulator: given
the agent's action $\tparam$, it constructs the active space
$\mathcal{A}(\tparam)$, runs \casci{} + sc-NEVPT2, and returns
a scalar reward.
Because the environment involves integer orbital selection followed
by a variational eigenvalue problem, it is non-differentiable with
respect to the action; gradients cannot be backpropagated through
the electronic-structure calculation.

\paragraph{Reward.}
The reward quantifies how well the selected active space reproduces
the \dmrg{} reference:
\begin{equation}
  r = -\bigl|\Edmrg - \Enevpt(\mathcal{A})\bigr|
      - \lambda_n \,|\mathcal{A}|,
  \label{eq:reward}
\end{equation}
where the first term penalizes deviation from the \dmrg{} reference
energy and the second term is a size regularizer that encourages
compact active spaces (fewer active orbitals); we set $\lambda_n = 0.05$.
Because $|\mathcal{A}|$ is a dimensionless orbital count, $\lambda_n$
carries the same units as energy, so that each additional active
orbital incurs a fixed energy-equivalent penalty balancing accuracy
against active-space compactness.

\paragraph{Policy optimization.}
The policy is optimized via proximal policy optimization
(PPO)~\cite{Schulman2017}, which is well suited to this setting
because it estimates policy gradients from sampled rewards without
requiring environment differentiability.
The clipped surrogate objective is
\begin{equation}
  \mathcal{L}^{\text{PPO}} =
    -\mathbb{E}\!\Bigl[
      \min\!\bigl(
        \rho_t \hat{A}_t,\;
        \text{clip}(\rho_t, 1\!-\!\epsilon, 1\!+\!\epsilon)\,\hat{A}_t
      \bigr)
    \Bigr]
    - \alpha_H\, H[\pi],
  \label{eq:ppo}
\end{equation}
where $\rho_t = \pi_{\phi,\text{new}}(\tilde{\tparam}_t \mid \mathbf{s}_t)
  /\pi_{\phi,\text{old}}(\tilde{\tparam}_t \mid \mathbf{s}_t)$ is the
importance sampling ratio computed from the unclipped Gaussian
proposal $\tilde{\tparam}_t$,
$\hat{A}_t = r_t - V_\psi(\mathbf{s}_t)$ is the advantage estimated
using a learned value baseline $V_\psi$ (a separate network with
the same architecture as the policy),
$\epsilon$ is the clip ratio, and
$\alpha_H H[\pi]$ is an entropy bonus that maintains exploration
and prevents premature collapse of the policy distribution.

\paragraph{Training loop.}
At each RL epoch the agent interacts with the environment for every
geometry in the training set: it observes the state $\mathbf{s}$,
samples a threshold $\tparam$ from the policy, runs the
\casci{} + sc-NEVPT2 calculation, receives a reward, and stores
the transition.
After collecting all transitions, the policy-network weights $\phi$,
log-standard-deviation $\log\sigma$, and value-network weights
$\psi$ are updated jointly via PPO over 200 gradient steps per
epoch, with separate Adam optimizers for the policy and value
networks and gradient norms clipped at 5.0.
Advantages are normalized to zero mean and unit variance for
stability.
The predictor weights $\theta$ are held fixed during this stage,
so the $\hat{s}_1$ network serves as a learned state representation
while the policy network $\pi_\phi$ learns to map each molecular
state to its optimal threshold.
Training converges in 6 RL epochs (each comprising a full pass over
all training geometries followed by the PPO update).

\subsection{Algorithm summary}
\label{sec:method:algo}

\Cref{alg:rlease} summarizes the full \rlease{} pipeline, from
data preparation through supervised $\hat{s}_1$ training,
PPO-based threshold optimization, and deployment.

\begin{algorithm}[H]
\caption{\rlease{} Training and Deployment}
\label{alg:rlease}
\begin{algorithmic}[1]
\Require Molecular geometries (XYZ), DMRG $s_1$ targets, DMRG reference
         energies, basis set
\Statex \textbf{Stage 1: Data preparation}
\State Compute HF orbitals for all geometries (with stability analysis)
\State Extract orbital descriptors $\{\mathbf{x}_i\}$
\Statex \textbf{Stage 2: Supervised $\hat{s}_1$ training}
\For{epoch $= 1, \dots, E_\text{sup}$}
  \State Forward pass: $\hat{s}_1 = f_\theta(\mathbf{x})$
  \State Compute Smooth-$L_1$ loss vs.\ DMRG $s_1$
  \State Update $\theta$ via AdamW
\EndFor
\Statex \textbf{Stage 3: RL threshold optimization}
\For{epoch $= 1, \dots, E_\text{RL}$}
  \For{each geometry in training set}
    \State Predict $\hat{s}_1$; compute state $\mathbf{s}$
    \State Sample $\tilde{\tparam} \sim \mathcal{N}\!\bigl(\mu_\phi(\mathbf{s}),\, \sigma^2\bigr)$; set $\tparam = \max(\tilde{\tparam},\, 0)$
    \State Build active space $\mathcal{A}(\tparam)$
    \State Compute $\Enevpt(\mathcal{A})$ via \casci{} + sc-NEVPT2
    \State Compute reward $r$ (\cref{eq:reward})
  \EndFor
  \State Compute advantages $\hat{A}_t = r_t - V_\psi(\mathbf{s}_t)$; normalize
  \State Update $(\phi, \log\sigma)$ via PPO (\cref{eq:ppo}); update $\psi$ via MSE on $r_t$
\EndFor
\Statex \textbf{Stage 4: Deployment}
\State Load trained $f_\theta$ and $\pi_\phi$
\State For new geometry: predict $\hat{s}_1$, compute $\tparam = \mu_\phi(\mathbf{s})$, build $\mathcal{A}(\tparam)$
\State Evaluate with one or more methods:
       CASCI + sc-NEVPT2, ASF-CCSD, ASF-CCSD(T)
\end{algorithmic}
\end{algorithm}

\subsection{NEVPT2 as the reward function}
\label{sec:method:reward}

At each RL step, \rlease{} constructs the active space
$\mathcal{A}(\tparam)$, classifies molecular orbitals into core,
active, and virtual sets, and runs a \casci{} calculation followed
by sc-NEVPT2 to obtain $\Enevpt$. 
No coupled-cluster calculations are required during training:
the RL loop uses only \hf{}, \casci{}, and \nevpt{}, making the
training pipeline fully multireference.

The NEVPT2 reward is evaluated on the fly for each proposed active space.
The number of active electrons is determined from the occupied
$\alpha$ and $\beta$ orbitals that remain after the inactive core has
been assigned.  This gives a CAS$(N_e,N_o)$ specification for each
candidate active space, where $N_o=|\mathcal{A}|$ and $N_e$ is the
corresponding number of active electrons.  Because the reward is computed
from the actual post-selection CASCI + NEVPT2 calculation, the learned
threshold is optimized directly for agreement with the DMRG reference
energy rather than for agreement with a hand-chosen active space.

\subsection{Multi-branch deployment: NEVPT2, ASF-CCSD, and ASF-CCSD(T)}
\label{sec:method:deploy}

A key advantage of RLEASE is that the learned active space is
\emph{method-agnostic}: although the threshold is trained using the
NEVPT2 reward, the resulting active-space selection can be used with
any method that requires an active space.
At inference time, we evaluate three downstream methods using
the same RLEASE-selected active space $\mathcal{A}$:
\begin{enumerate}
  \item \textbf{CASCI + sc-NEVPT2}: the purely multireference
        pathway used during training (\cref{eq:nevpt2}).
  \item \textbf{ASF-CCSD}: the additive-subtractive composite
        correction applied to full-space CCSD (\cref{eq:asf}).
  \item \textbf{ASF-CCSD(T)}: the same correction applied to
        full-space CCSD(T).
\end{enumerate}
This multi-branch strategy requires coupled-cluster calculations
only at deployment time, not during the RL training loop.
The NEVPT2 reward serves as a computationally affordable proxy
that drives the RL agent to select active spaces capturing the
dominant static correlation; these same active spaces then
improve the coupled-cluster description via the ASF correction.

For molecules where single-reference CCSD or CCSD(T) already
provides a good description, the ASF correction is small and the
ASF energy remains close to the full-space CC result.
For strongly correlated systems where CC breaks down, the CASCI
replacement of the internal CC contribution corrects the largest
errors.  The NEVPT2 pathway provides a fully multireference
alternative that does not rely on the CC framework at all.

\FloatBarrier
\section{Computational Details}
\label{sec:comp}

\subsection{Molecular dataset}
\label{sec:comp:dataset}

\rlease{} is trained on only \textbf{three} molecules
(Na$_2$, ClF, and SiO$_2$) sampled along their potential-energy surfaces, together with calculations on the corresponding constituent atoms (Na, Cl, F, Si, and O). It is then deployed without retraining to the test molecules considered in this work:
\begin{itemize}
  \item \textbf{Main-group diatomics:}
        N$_2$, P$_2$, LiH, NaH, HBr.
  \item \textbf{An open-shell radical:} FO.
  \item \textbf{Polyatomic molecules:}
        BeH$_2$, BH$_3$, CH$_4$, NH$_3$.
  \item \textbf{3$d$ transition-metal hydrides:}
        ZnH, CuH.
\end{itemize}

Each molecule is represented by 7 to 41 geometries along a
bond-stretching coordinate, from near-equilibrium to dissociation
(typically $\sim$30 points per species). The training data are generated in the minimal STO-3G basis, whereas all test-set predictions reported below are performed in the larger cc-pVDZ basis.
This train/test split is deliberately extreme: by training on a
minimal set of molecules, we test whether \rlease{} learns
transferable chemical intuition for active-space selection rather
than memorizing molecule-specific patterns.

\subsection{Reference calculations}
\label{sec:comp:reference}

All Hartree--Fock calculations use the PySCF
package~\cite{PySCF2018}.
Open-shell systems are treated with unrestricted HF (UHF) with
SCF stability analysis to ensure convergence to the lowest-energy
solution.
For UHF references, the active-space mask is constructed from
the $\alpha$-spin canonical orbital ordering.
The $\alpha$-spin MO coefficients are then used to build an
ROHF-based restricted CASCI reference; sc-NEVPT2 is applied to this
restricted CASCI wavefunction via PySCF's \texttt{mrpt.NEVPT}
interface.
This pathway requires symmetric $\alpha$/$\beta$ core partitions (i.e., the same number of inactive orbitals for each spin).
All open-shell geometries reported in the PES curves satisfied
this symmetric-core condition.
Descriptors are extracted from the same canonical UHF
$\alpha$-spin orbitals, ensuring consistency between the
active-space selection and the energy evaluation.
%
We employ CASCI rather than CASSCF throughout this work: the
Hartree--Fock orbitals are used without further orbital optimization
in the active space.
This choice avoids the additional cost and potential convergence
difficulties of the CASSCF orbital-optimization step, and ensures
that the orbital basis is consistent between the descriptor
extraction (which uses HF orbitals) and the downstream energy
evaluation.
Incorporating CASSCF orbital optimization is a natural extension
that may improve results for systems with strong orbital relaxation
effects.

All \dmrg{} calculations are performed with the
\textsc{block2} package~\cite{Zhai2021,Zhai2023} using
$S_z$ symmetry and full-space active spaces (all MOs active).
The sweep schedule ramps the bond dimension from $D/2$ to the
target $D$ over 16 sweeps with a noise tapering of
$10^{-4} \to 10^{-5} \to 0$ and convergence thresholds of
$10^{-8}$ (ramp) and $10^{-10}$ (final sweeps).

For \textbf{training}, converged \dmrg{} energies and single-orbital entropies
$s_1$ are computed in the STO-3G basis with bond dimension $D = 1500$.
The smaller basis keeps the training cost manageable while providing
physically meaningful $s_1$ targets.
For \textbf{inference}, the trained model is deployed
in the cc-pVDZ basis.  The RLEASE-selected active space is
evaluated with three methods: CASCI + sc-NEVPT2, ASF-CCSD, and
ASF-CCSD(T).  All coupled-cluster calculations (full-space and
internal-only) are performed with PySCF's CCSD and CCSD(T)
implementations~\cite{PySCF2018}.
\dmrg{} reference energies are obtained at $D = 1500$
in cc-pVDZ for benchmarking all three methods.

$s_1$ values are extracted from the \dmrg{} one-orbital reduced
density matrices and serve as both training targets for the neural
network and as the ground truth for evaluating orbital selection
quality.
\clearpage
\FloatBarrier
\subsection{RLEASE hyperparameters}
\label{sec:comp:hyper}

The key hyperparameters used for the supervised $\hat{s}_1$
predictor, PPO-based threshold optimization, and the reference
DMRG calculations are summarized in \cref{tab:hyperparams}.

\begin{table}[h]
\centering
\caption{Key hyperparameters used in \rlease{} training.}
\label{tab:hyperparams}
\begin{tabular}{lc}
\toprule
\textbf{Parameter} & \textbf{Value} \\
\midrule
\multicolumn{2}{l}{\textit{$\hat{s}_1$ predictor}} \\
Hidden dimension $H$        & 256 \\
Depth $L$                   & 8 \\
Dropout                     & 0.0 \\
Learning rate                & $5 \times 10^{-3}$ \\
Weight decay                 & $1 \times 10^{-4}$ \\
Epochs (supervised)          & 2000 \\
Early stopping patience      & 1850 \\
Batch size                   & 4 \\
Loss function                & Smooth-$L_1$ ($\beta=0.05$) \\
Optimizer                    & AdamW \\
LR schedule                  & Cosine annealing \\
Gradient clip norm           & 5.0 \\
Target transform             & $\log(1+x)$  normalization \\
\midrule
\multicolumn{2}{l}{\textit{RL threshold}} \\
RL epochs                    & 6 \\
PPO clip $\epsilon$          & 0.3 \\
PPO update steps per epoch    & 200 \\
RL learning rate             & $5 \times 10^{-3}$ \\
$\sigma_0$                   & 0.02 \\
$\tau_0$ (initial threshold) & 0.1 \\
$\lambda_n$ & 0.05\\
\midrule
\multicolumn{2}{l}{\textit{DMRG}} \\
Bond dimension $D$ (train)   & 1500 \\
Bond dimension $D$ (eval)    & 1500 \\
Number of sweeps             & 16 \\
Symmetry                     & $S_z$ \\
Solver                       & \textsc{block2} \\
\midrule
\multicolumn{2}{l}{\textit{General}} \\
Training basis set           & STO-3G \\
Open-shell method            & UHF \\
Orbital ordering             & Canonical \\
Stability check              & Yes \\
\bottomrule
\end{tabular}
\end{table}

\FloatBarrier
\section{Results and Discussion}
\label{sec:results}

\subsection{Accuracy of the \texorpdfstring{$\hat{s}_1$}{s1} predictor}
\label{sec:results:s1}

The supervised $\hat{s}_1$ predictor is trained on orbital data
from the three training molecules and their PES geometries
(STO-3G basis, $\sim$30 geometries per molecule along the binding
curve).
The orbital-level data from all molecules and geometries are pooled
and randomly split 70:30 into training and test sets.
On the held-out test set, the predictor achieves
$R^2 = 0.99$, RMSE $= 0.071$, and MAE $= 0.054$,
indicating that the neural network accurately reproduces the
\dmrg{}-derived $\sone$ values from Hartree--Fock
descriptors alone.

We note that the quality of $\hat{s}_1$ prediction is a necessary
but not sufficient condition for good active-space selection: the RL
threshold can partially compensate for prediction errors by shifting
the decision boundary.
The key metric is therefore the downstream energy accuracy
(\cref{sec:results:pes}), not the orbital-level prediction error
in isolation.

\subsection{Active-space selection quality}
\label{sec:results:selection}

We compare the active spaces selected by \rlease{} against three
reference methods on the same set of molecules and geometries:
autoCAS~\cite{Stein2016}, which applies a plateau-based threshold
to $\sone$ values from a pilot \dmrg{} calculation; and two fixed
$\sone$ thresholds, $\tparam = 0.1$ and $\tparam = 0.1\ln 4$, applied
directly to the \dmrg{}-derived $\sone$ values.
The latter two represent the simplest possible entropy-thresholding
baselines, while autoCAS represents the current state-of-the-art
automated selection method.

\begin{figure}[ht]
  \centering
  \includegraphics[width=0.49\textwidth]{%
    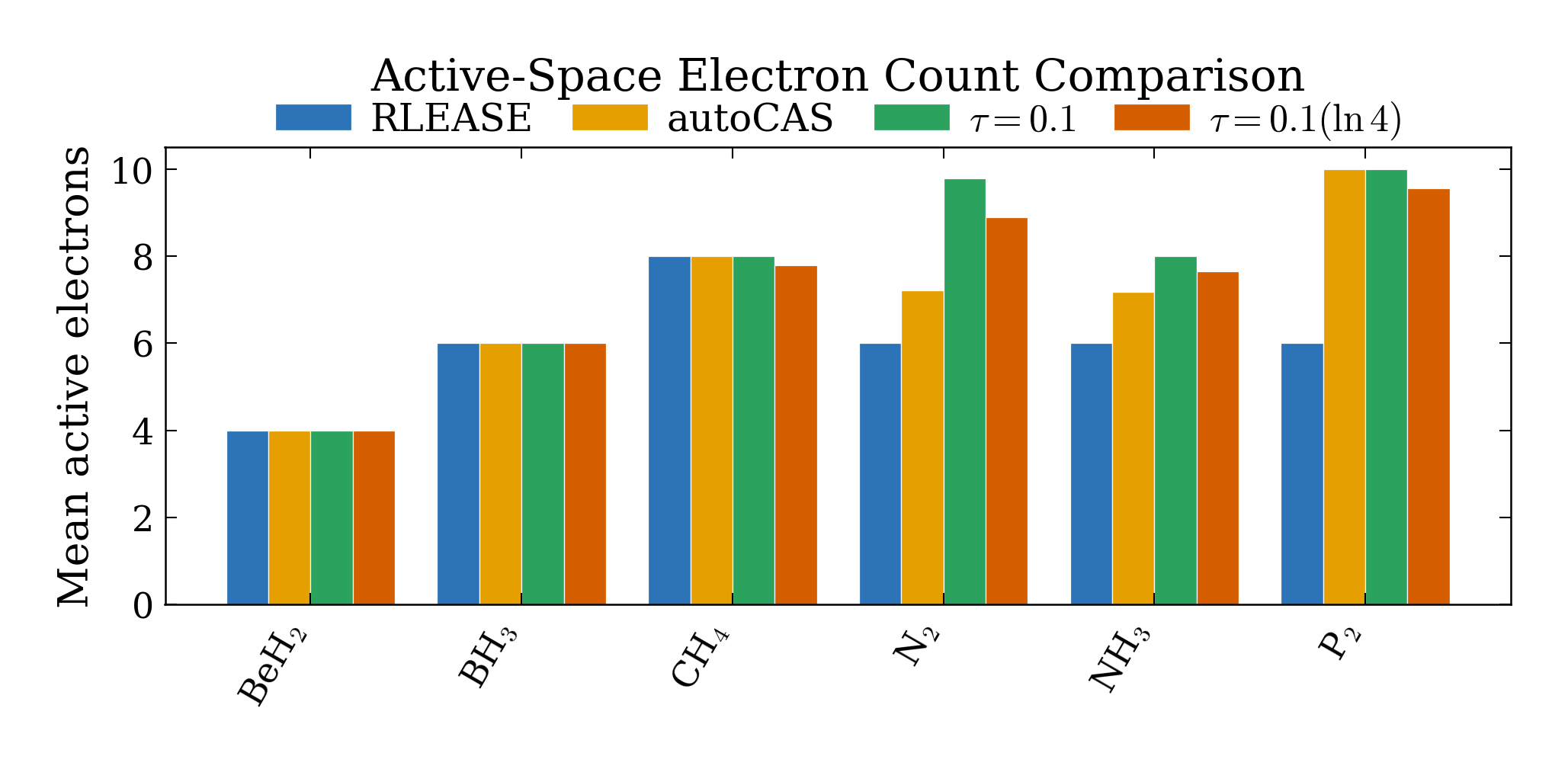}\hfill
  \includegraphics[width=0.49\textwidth]{%
    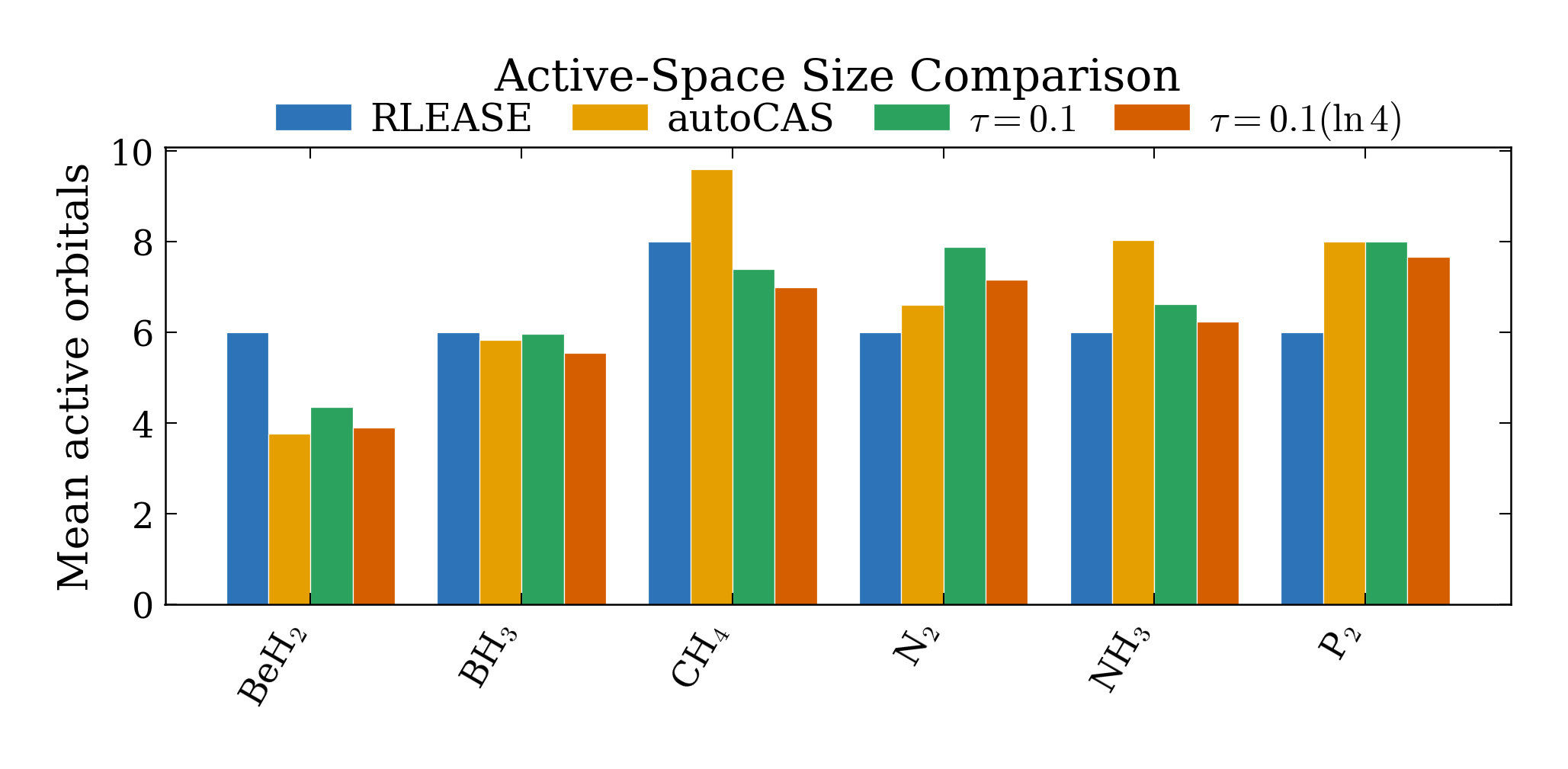}
  \caption{%
    Per-molecule active-space sizes selected by \rlease{}, autoCAS,
    $\tparam=0.1$, and $\tparam=0.1\ln 4$, averaged over all
    geometries along each PES, following the CAS($N_e$,$N_o$)
    convention.
    \textbf{Left:} number of active electrons $N_e$.
    \textbf{Right:} number of active orbitals $N_o$.
    \rlease{} consistently selects compact active spaces (typically
    4--8 orbitals for main-group species) while maintaining high
    orbital overlap with the reference methods (see \cref{fig:jaccard}).
  }
  \label{fig:active_sizes}
\end{figure}

\begin{figure}[ht]
  \centering
  \includegraphics[width=\textwidth]{%
    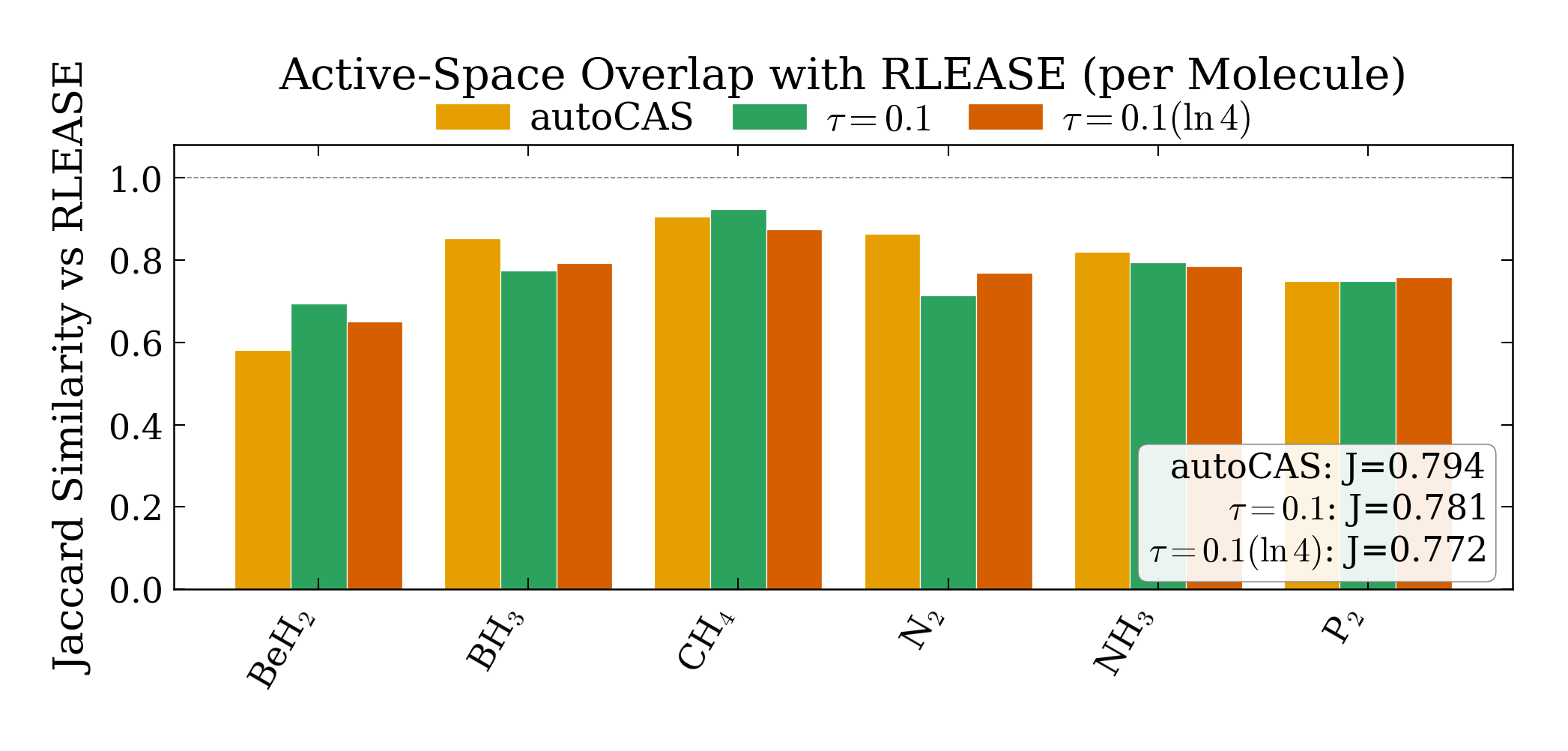}
  \caption{%
    Per-molecule Jaccard similarity between \rlease{}-selected active
    spaces and those of autoCAS, $\tparam = 0.1$, and
    $\tparam = 0.1\ln 4$, averaged over all geometries.
    A Jaccard of 1.0 indicates identical active-space orbital sets.
    \rlease{} achieves strong agreement with autoCAS and both fixed
    thresholds across the majority of molecules, with the largest
    discrepancies arising for molecules with ambiguous plateau
    structure in the $\sone$ profile.
  }
  \label{fig:jaccard}
\end{figure}

\begin{figure}[ht]
  \centering
  \includegraphics[width=1\textwidth]{%
    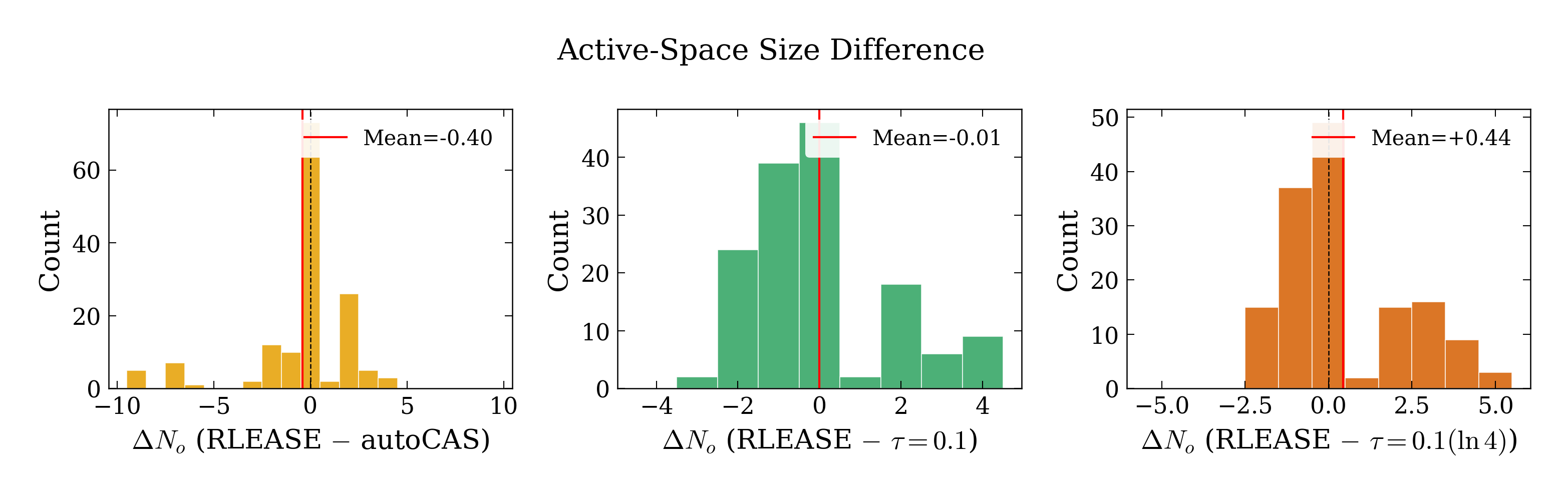}
  \caption{%
    Distribution of \emph{per-geometry} active-orbital count
    differences
    $\Delta N_o = N_o^\text{RLEASE} - N_o^\text{ref}$ for each
    reference method.
    Each count in the histogram corresponds to a single geometry of
    a single molecule (BeH$_2$, BH$_3$, CH$_4$, N$_2$, NH$_3$,
    P$_2$); these are \emph{not} molecule-averaged values.
    The distributions are approximately centered near zero,
    indicating that \rlease{} selects active spaces of comparable
    size to the reference methods overall.
    The tails (e.g.\ $\Delta N_o \approx -7$ to $-9$) arise from
    isolated geometries of CH$_4$ and NH$_3$ compared to autoCAS,
    where autoCAS selects substantially larger active spaces
    (up to 17 and 13 orbitals, respectively) at certain
    stretched-bond configurations, while \rlease{} maintains
    more compact selections.
  }
  \label{fig:size_diff}
\end{figure}

The active-space sizes selected by each method are compared in
\cref{fig:active_sizes}, which shows both the number of active orbitals
and active electrons per molecule.
\rlease{} selects compact active spaces that are broadly consistent
with those of autoCAS and the fixed-threshold baselines.
\Cref{fig:jaccard} shows the per-molecule Jaccard similarity between
\rlease{} and each reference method.
The majority of molecules exhibit Jaccard values above 0.7, indicating
substantial orbital-level agreement despite the methods using very
different selection criteria: autoCAS uses \dmrg{}-derived $\sone$
values with an empirical plateau algorithm, while \rlease{} predicts
$\hat{s}_1$ from HF descriptors and optimizes the threshold via RL.
BeH$_2$ shows the weakest agreement, particularly against autoCAS,
reflecting ambiguity in the $\sone$ plateau structure for this
system.
\Cref{fig:size_diff} shows the distribution of per-geometry
active-space size differences; each entry is a single geometry of a
single molecule, not a molecule average.
The distributions are approximately centered near zero,
confirming that \rlease{} does not systematically under- or
over-select relative to the reference methods.
The tails at large negative $\Delta N_o$ arise specifically from
CH$_4$ and NH$_3$ geometries compared to autoCAS, where autoCAS
selects up to 17 and 13 orbitals, respectively, while \rlease{}
selects 8 and 6.

\FloatBarrier
\subsection{Potential energy surfaces}
\label{sec:results:pes}

\begin{figure}[ht]
  \centering
  \includegraphics[width=0.48\textwidth]{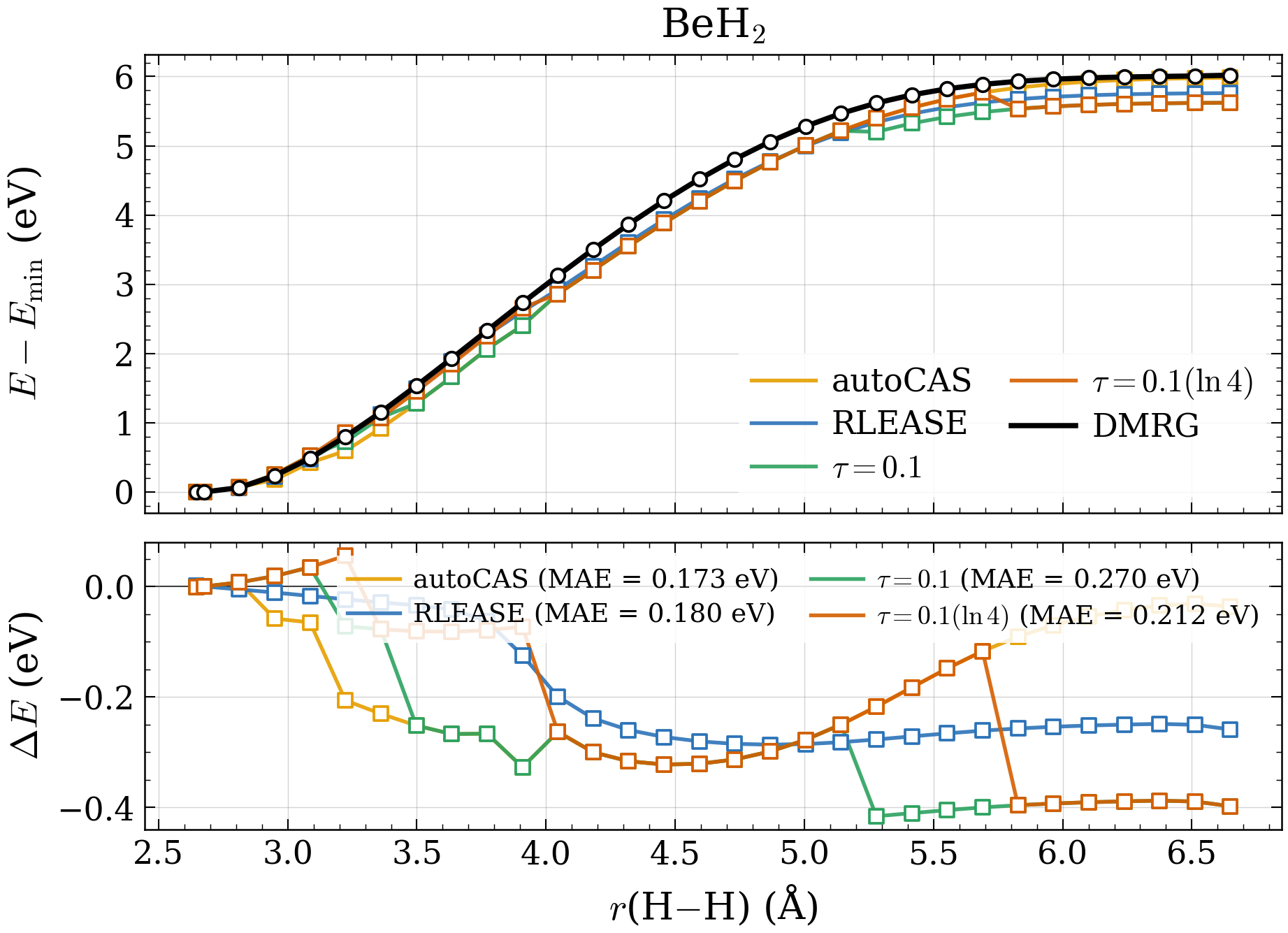}\hfill
  \includegraphics[width=0.48\textwidth]{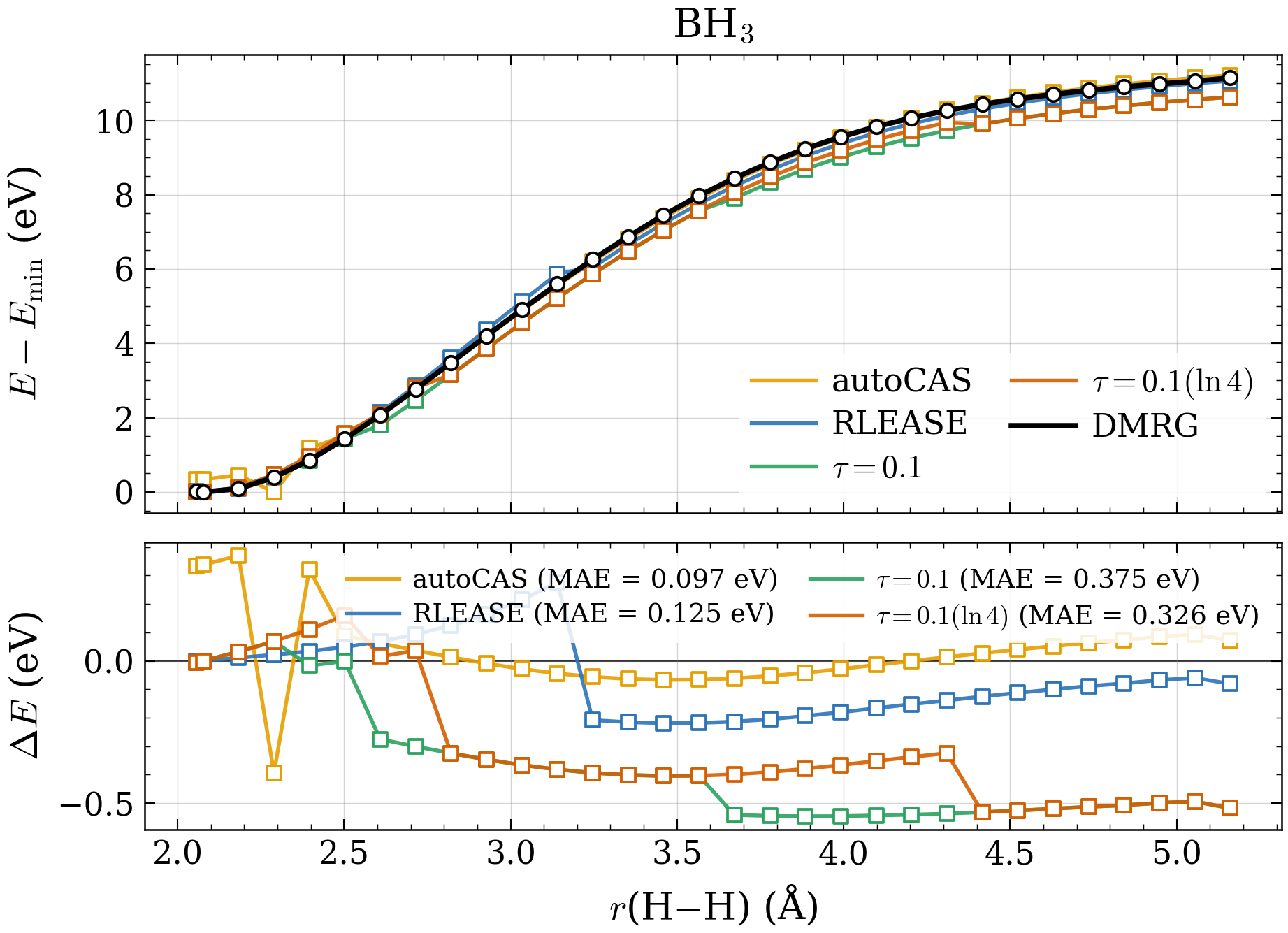}\\[4pt]
  \includegraphics[width=0.48\textwidth]{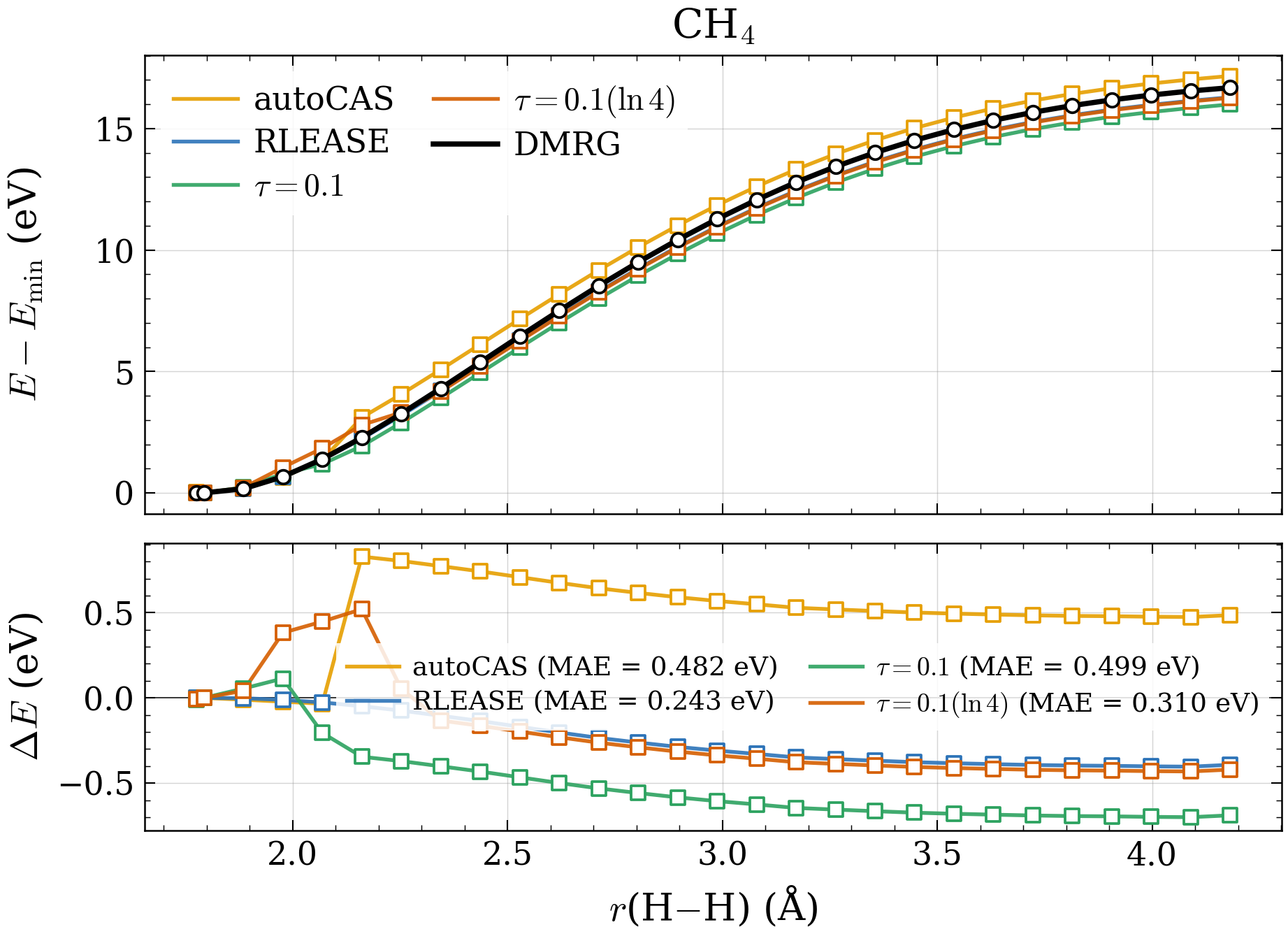}\hfill
  \includegraphics[width=0.48\textwidth]{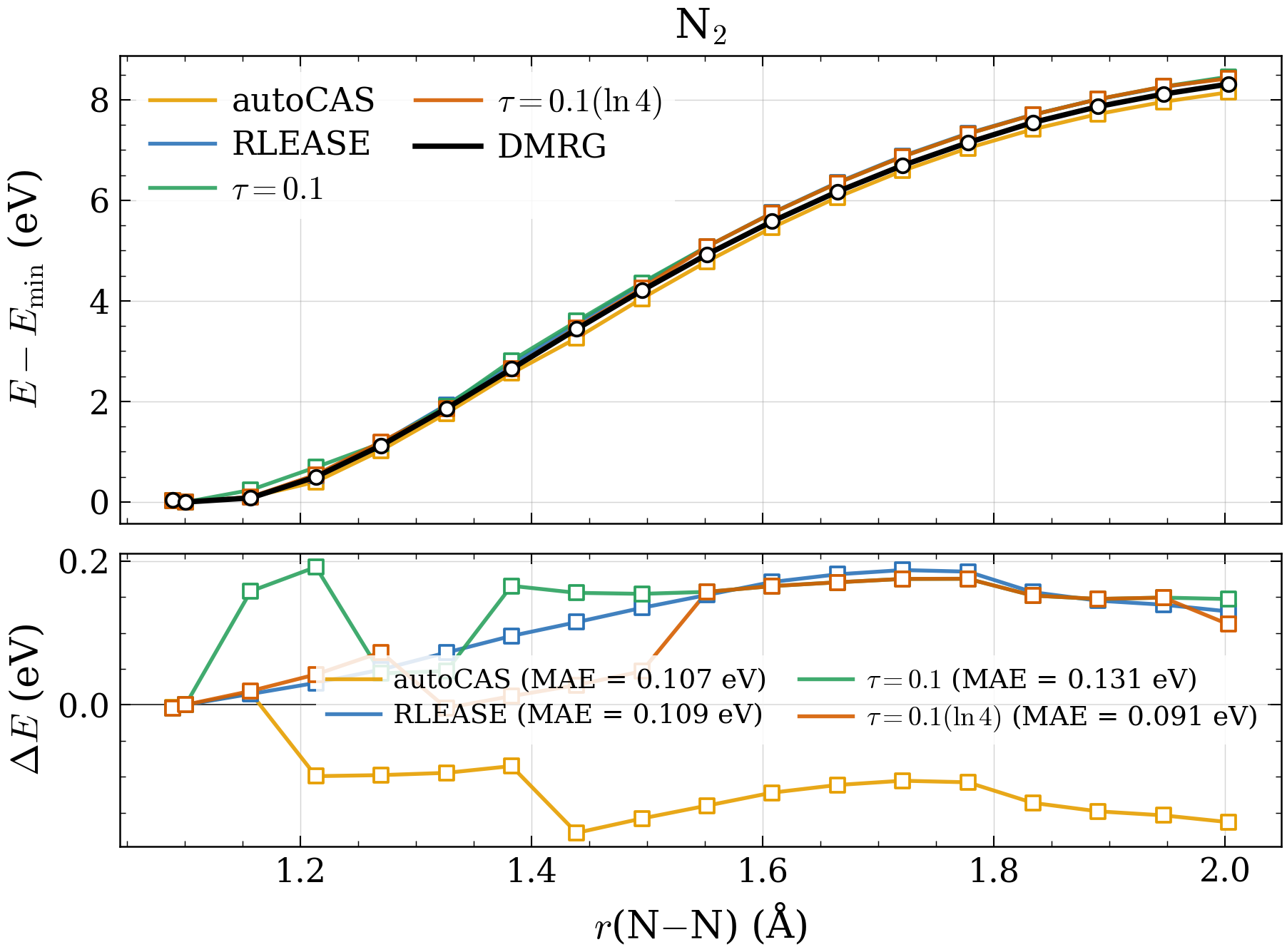}\\[4pt]
  \includegraphics[width=0.48\textwidth]{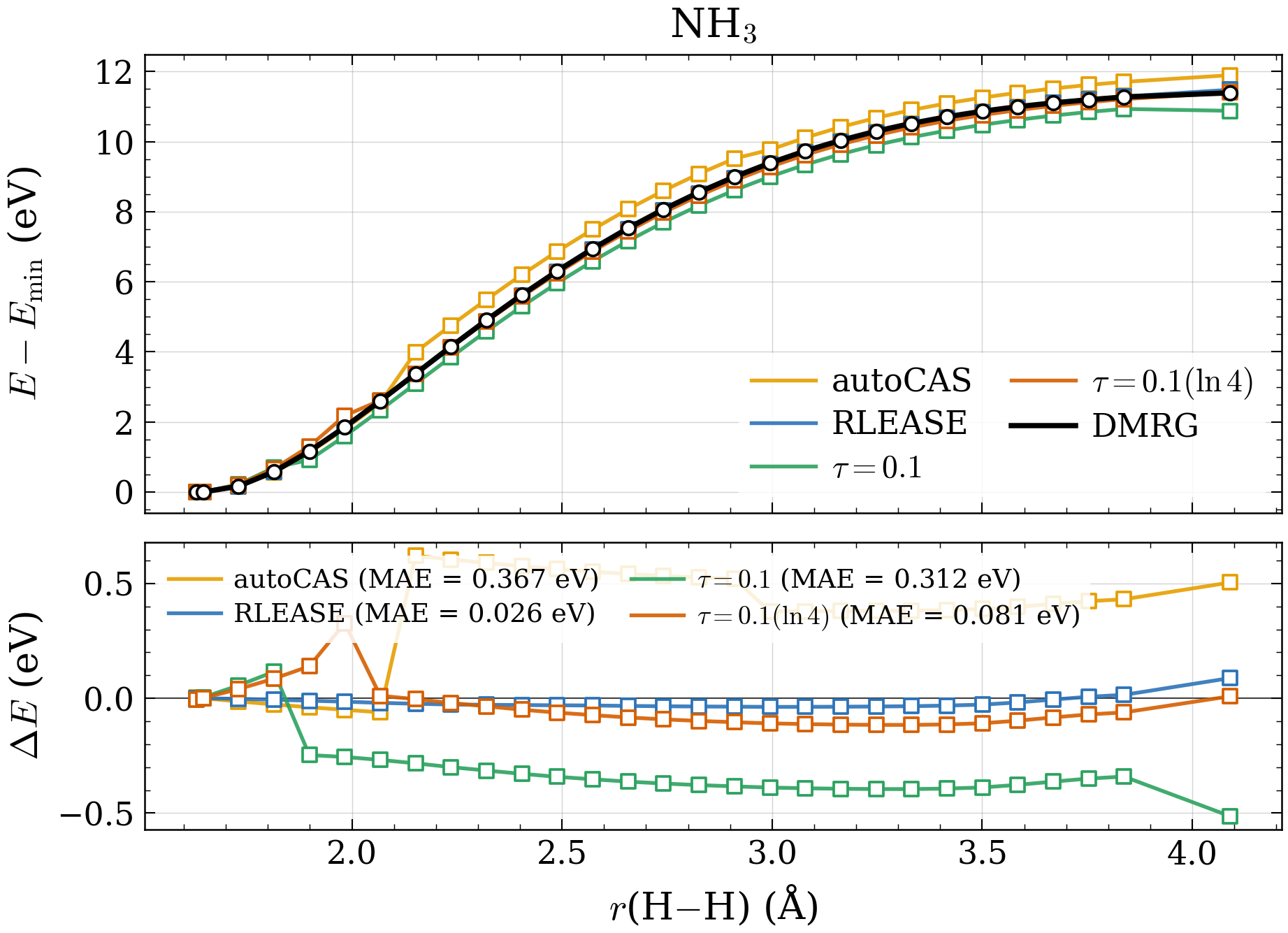}\hfill
  \includegraphics[width=0.48\textwidth]{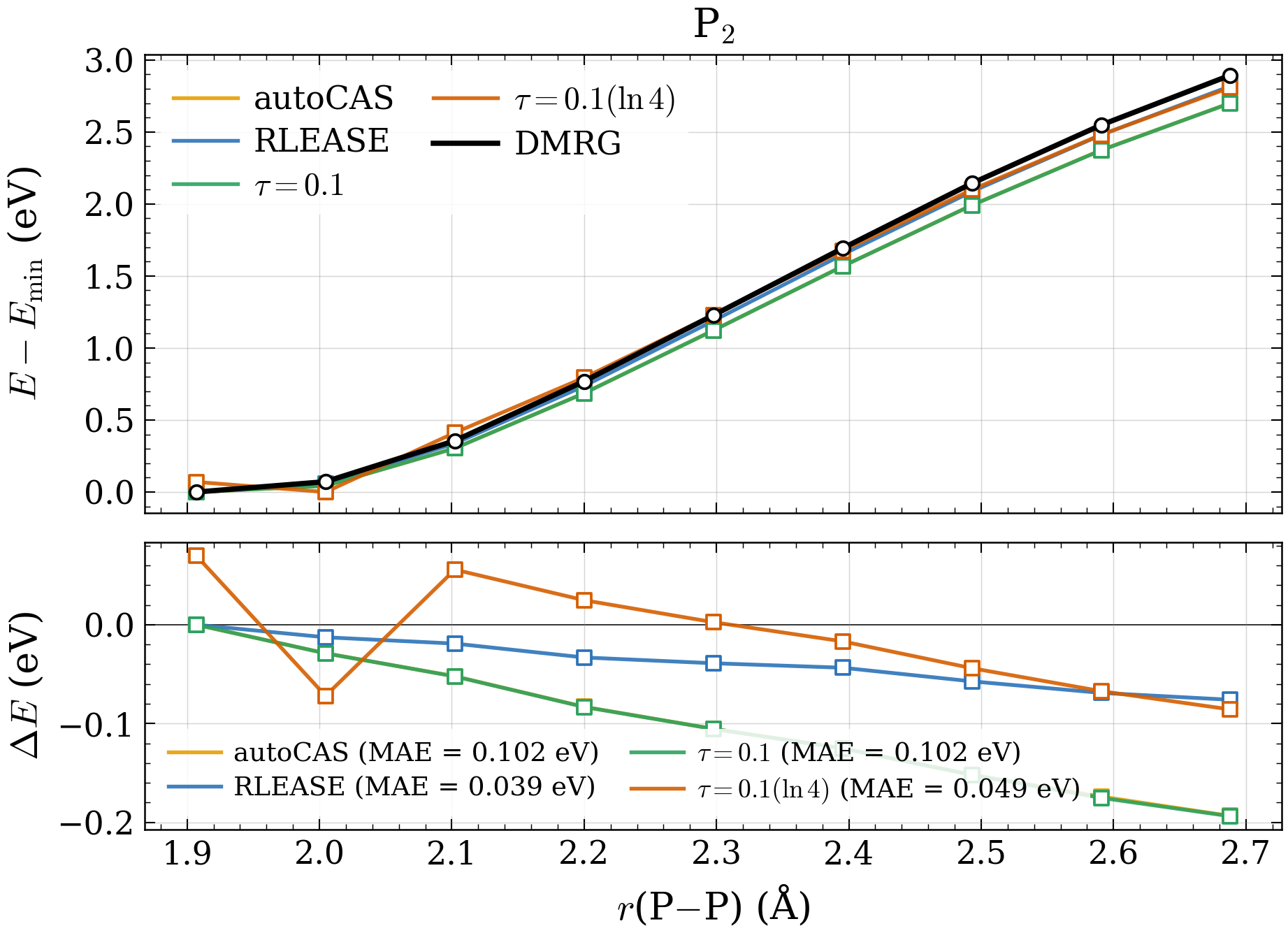}
  \caption{%
    Relative PES binding curves for CASCI + \nevpt{} using active spaces
    selected by \rlease{}, autoCAS, $\tparam=0.1$, and $\tparam=0.1\ln 4$,
    compared to \dmrg{} ($D=1500$, cc-pVDZ) reference energies (black).
    Energies are plotted relative to the minimum along each curve.
  }
  \label{fig:pes_nevpt2}
\end{figure}

\begin{figure}[ht]
  \centering
  \includegraphics[width=0.48\textwidth]{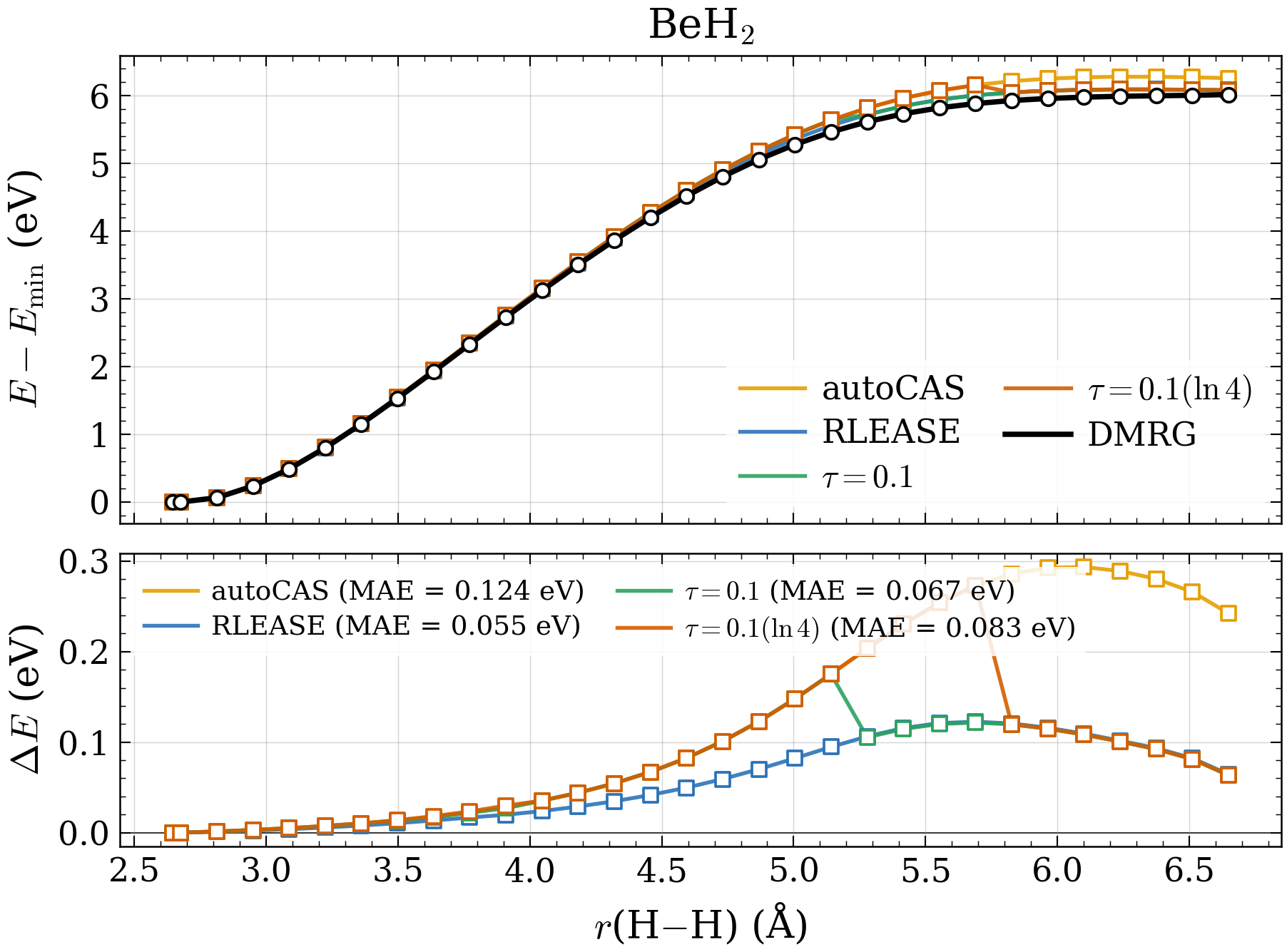}\hfill
  \includegraphics[width=0.48\textwidth]{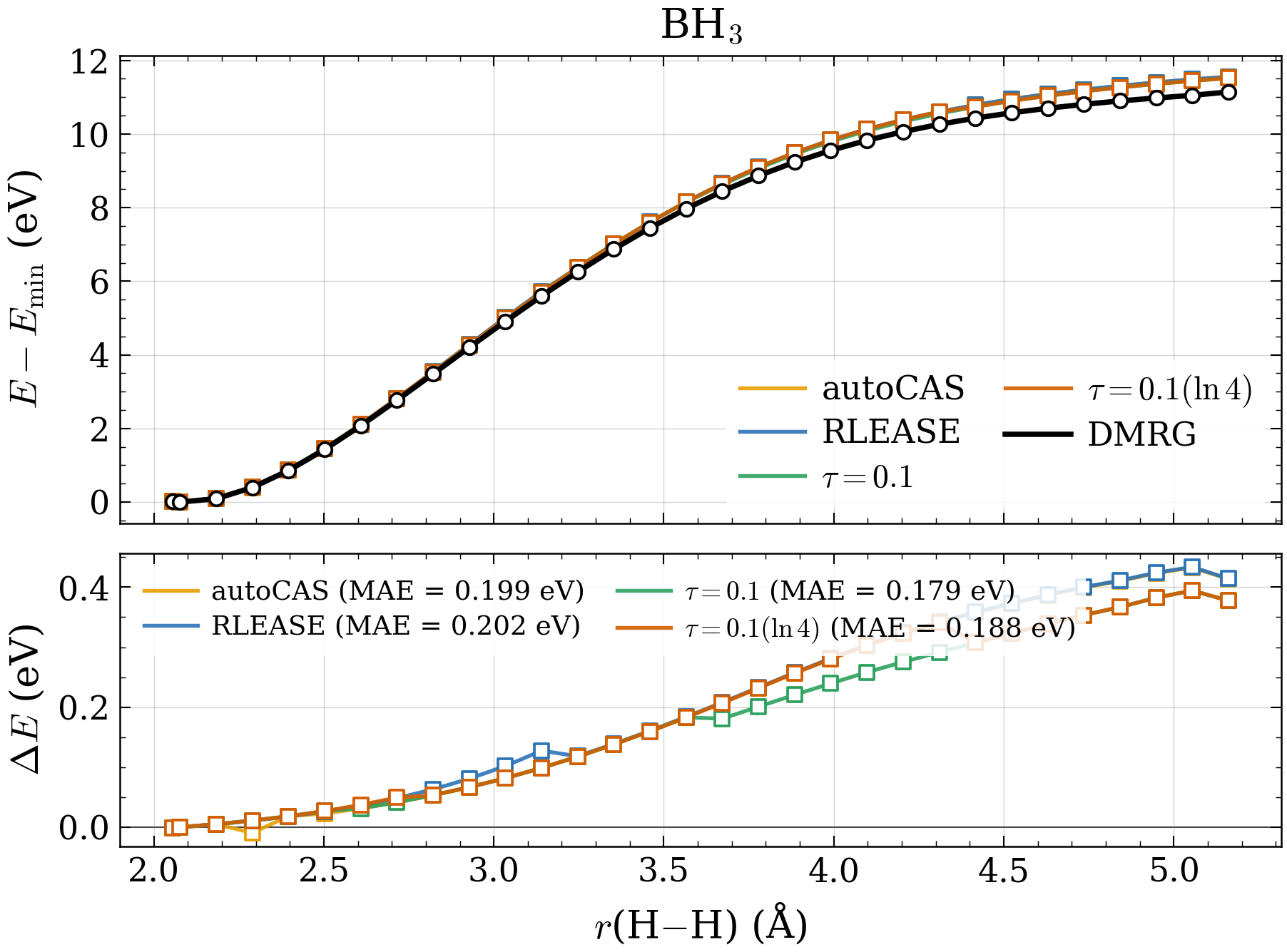}\\[4pt]
  \includegraphics[width=0.48\textwidth]{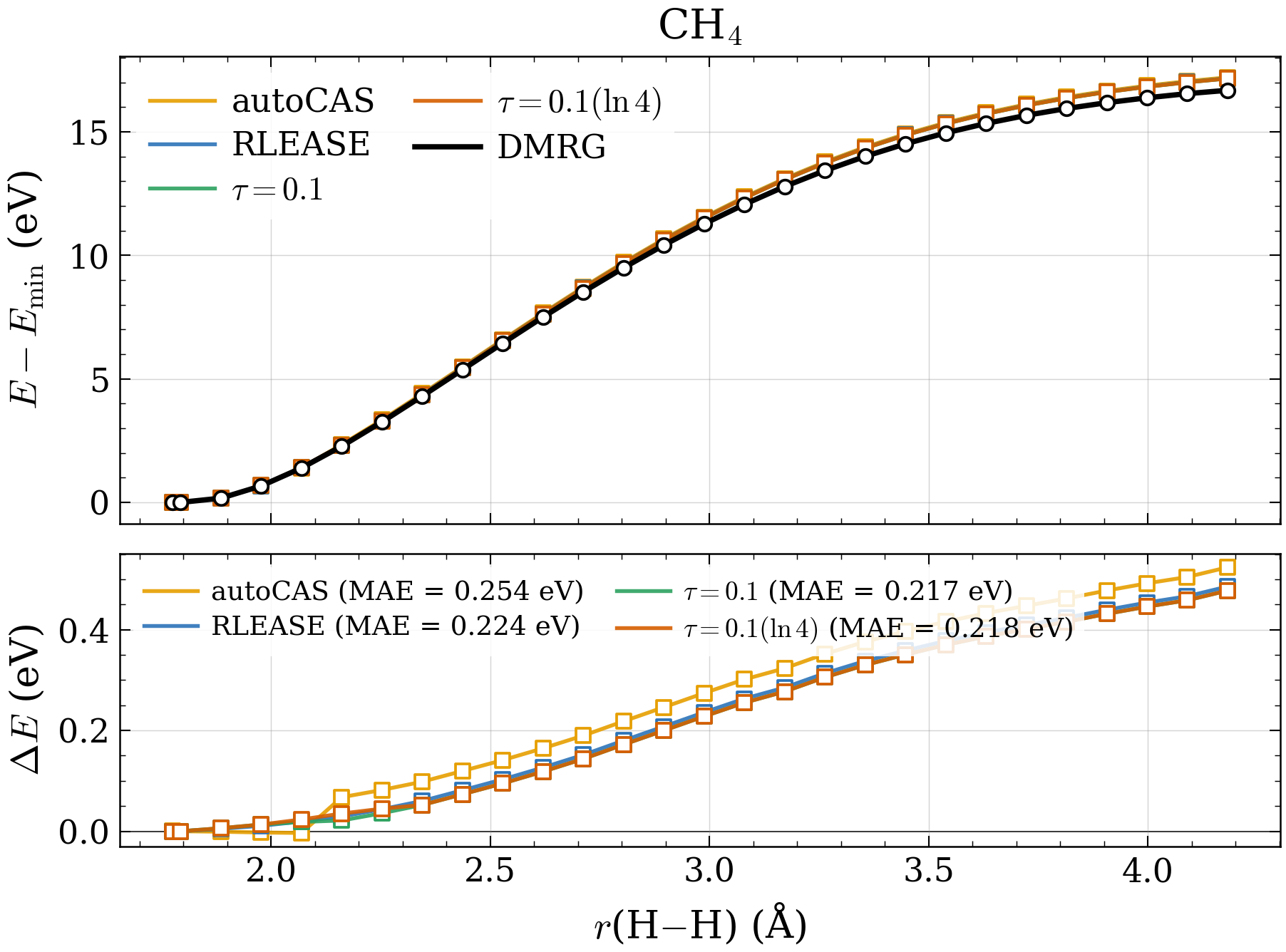}\hfill
  \includegraphics[width=0.48\textwidth]{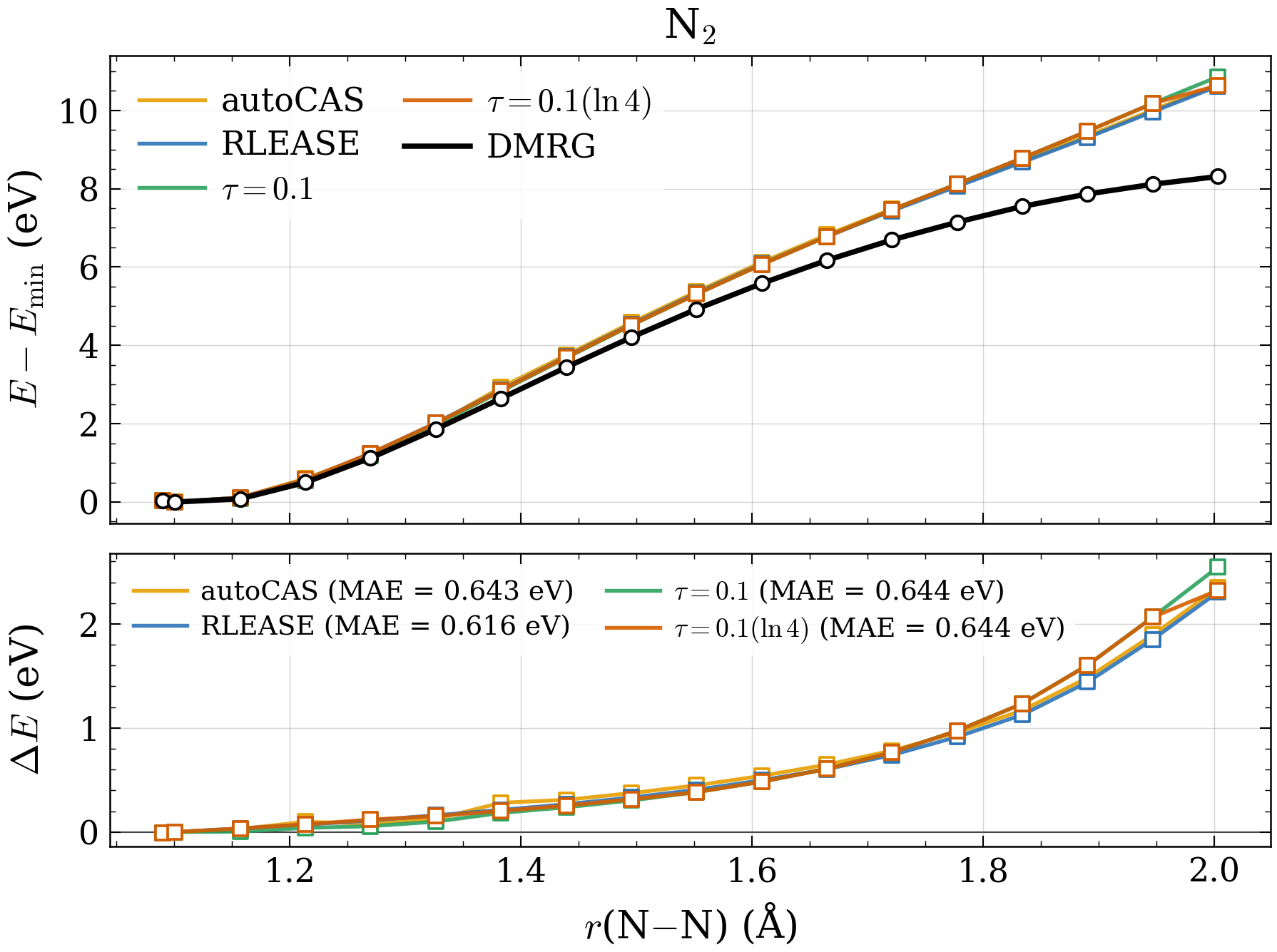}\\[4pt]
  \includegraphics[width=0.48\textwidth]{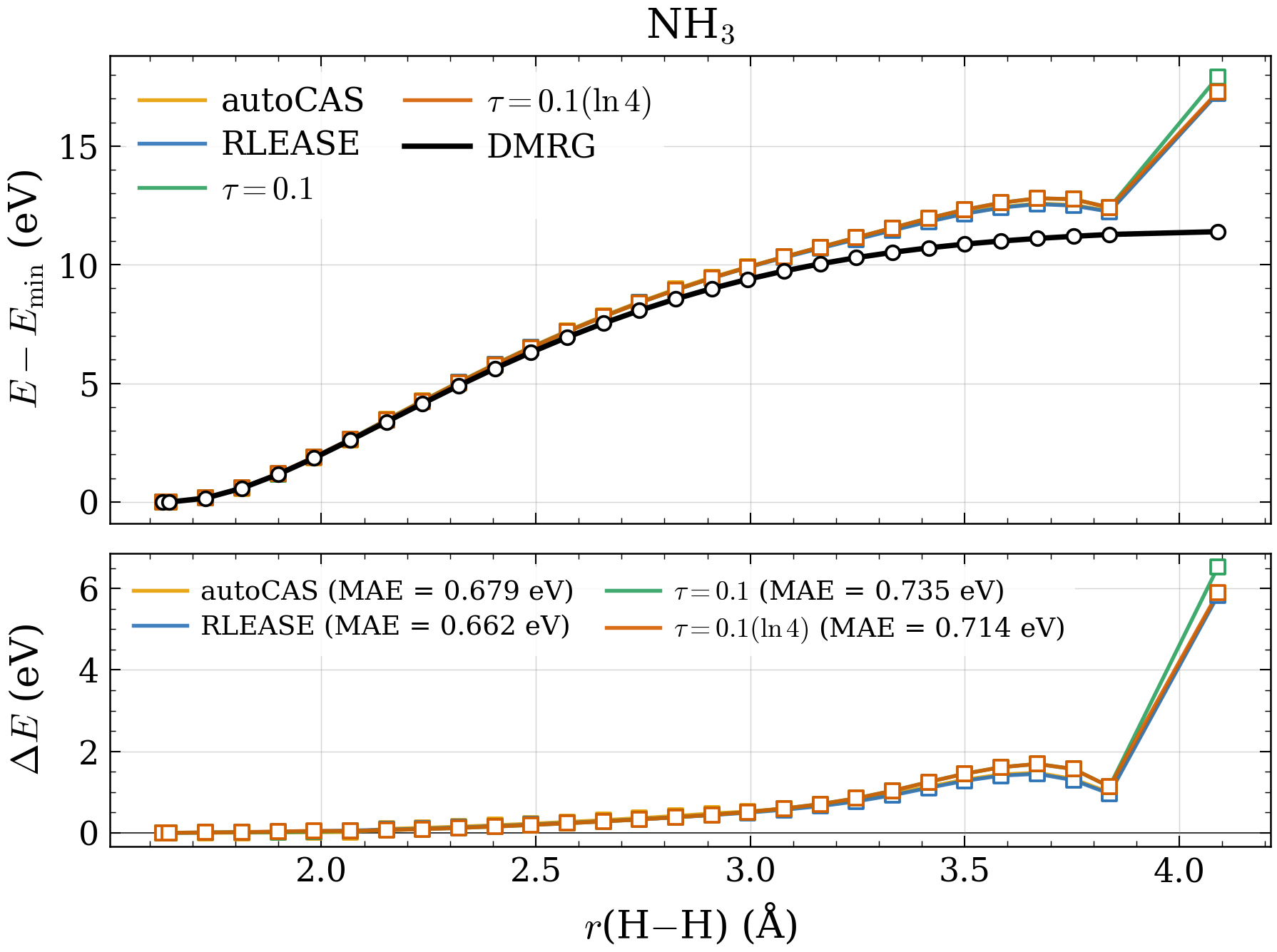}\hfill
  \includegraphics[width=0.48\textwidth]{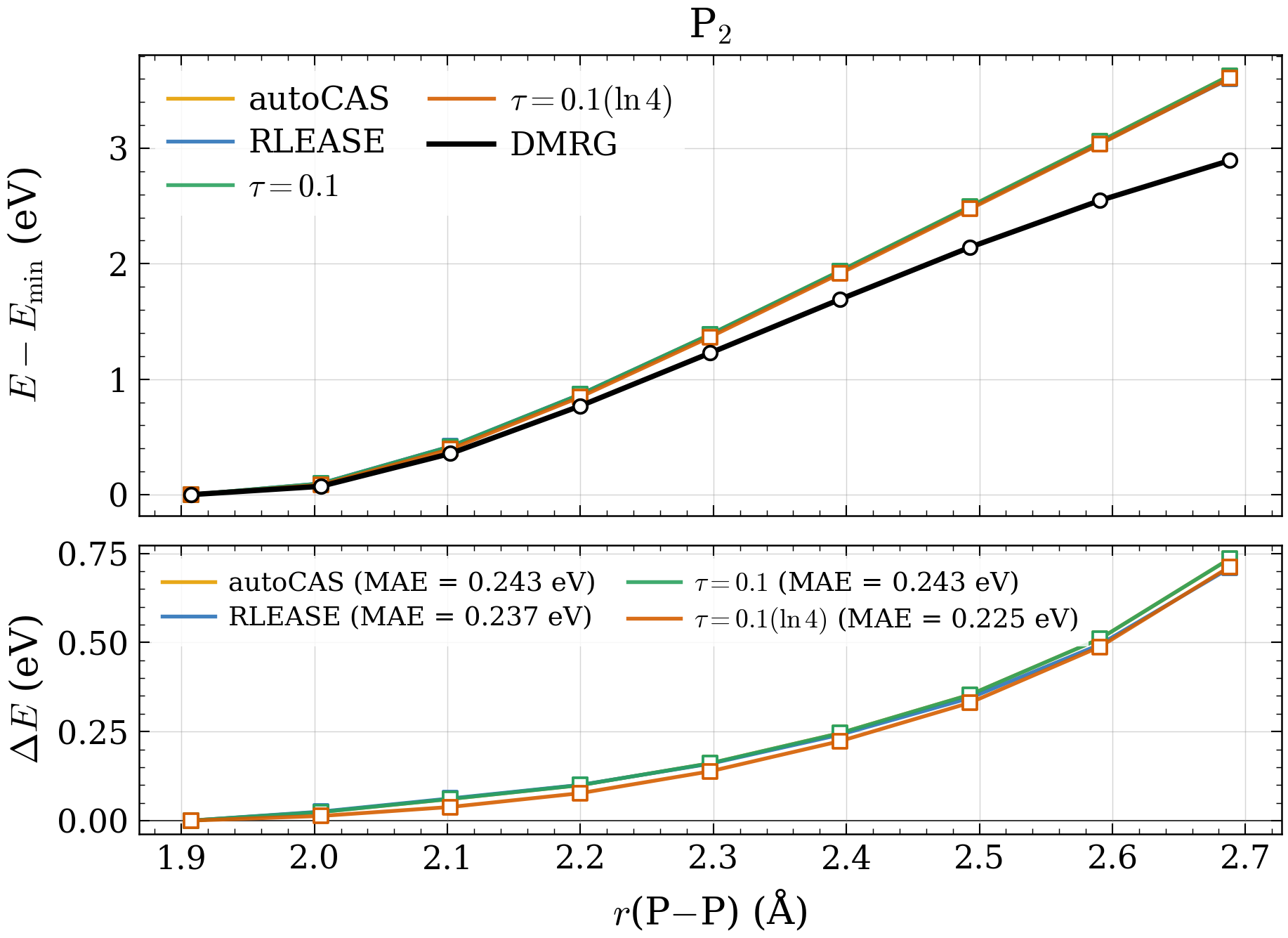}
  \caption{%
    Relative PES binding curves for ASF-CCSD using active spaces
    selected by \rlease{}, autoCAS, $\tparam=0.1$, and
    $\tparam=0.1\ln 4$, compared to \dmrg{} ($D=1500$, cc-pVDZ)
    reference energies (black).
    The ASF correction replaces the coupled-cluster description of
    the active orbitals with CASCI, recovering static correlation
    while retaining the full-space CCSD treatment of dynamic
    correlation.
  }
  \label{fig:pes_asfccsd}
\end{figure}

\begin{figure}[ht]
  \centering
  \includegraphics[width=0.48\textwidth]{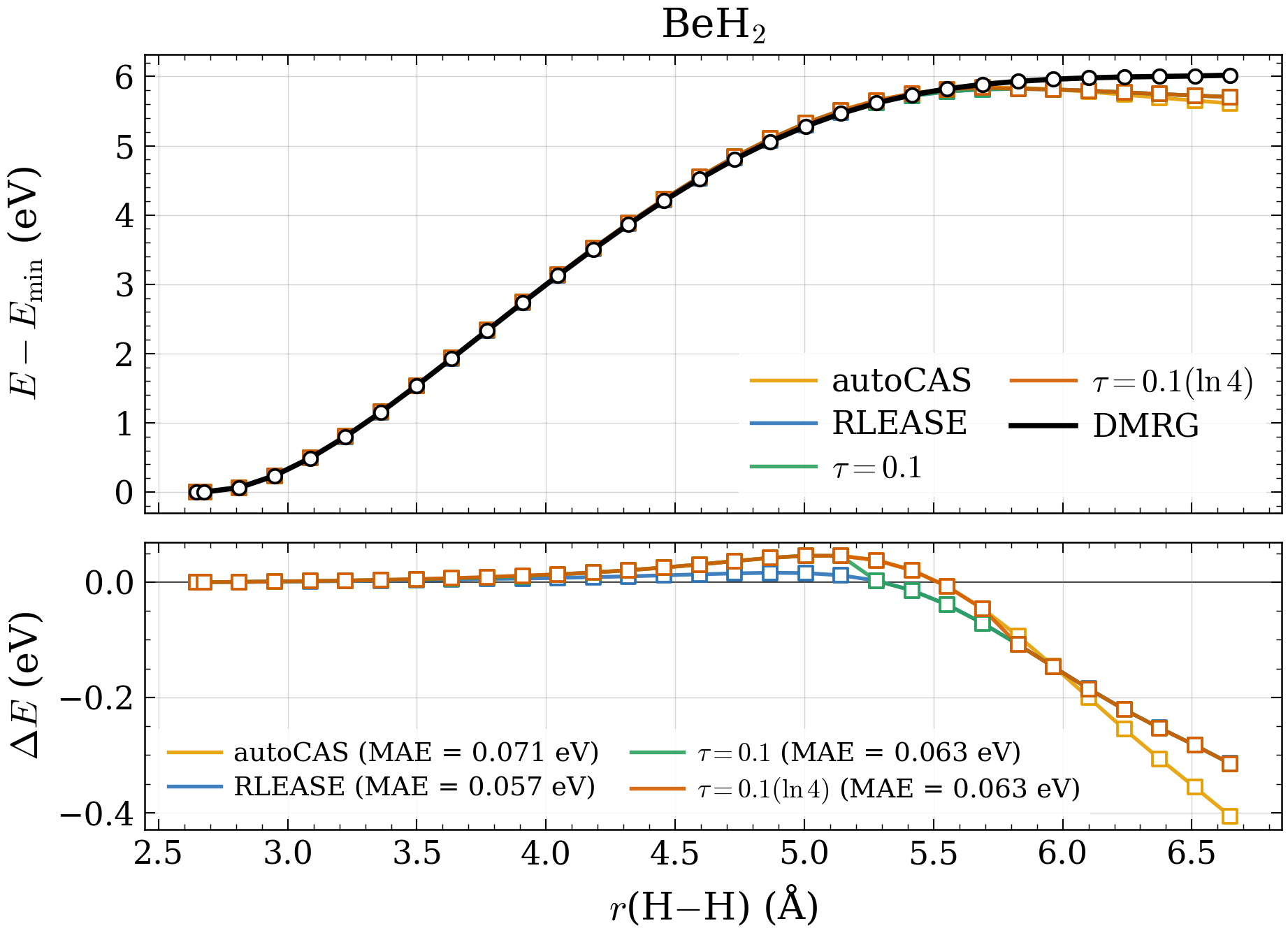}\hfill
  \includegraphics[width=0.48\textwidth]{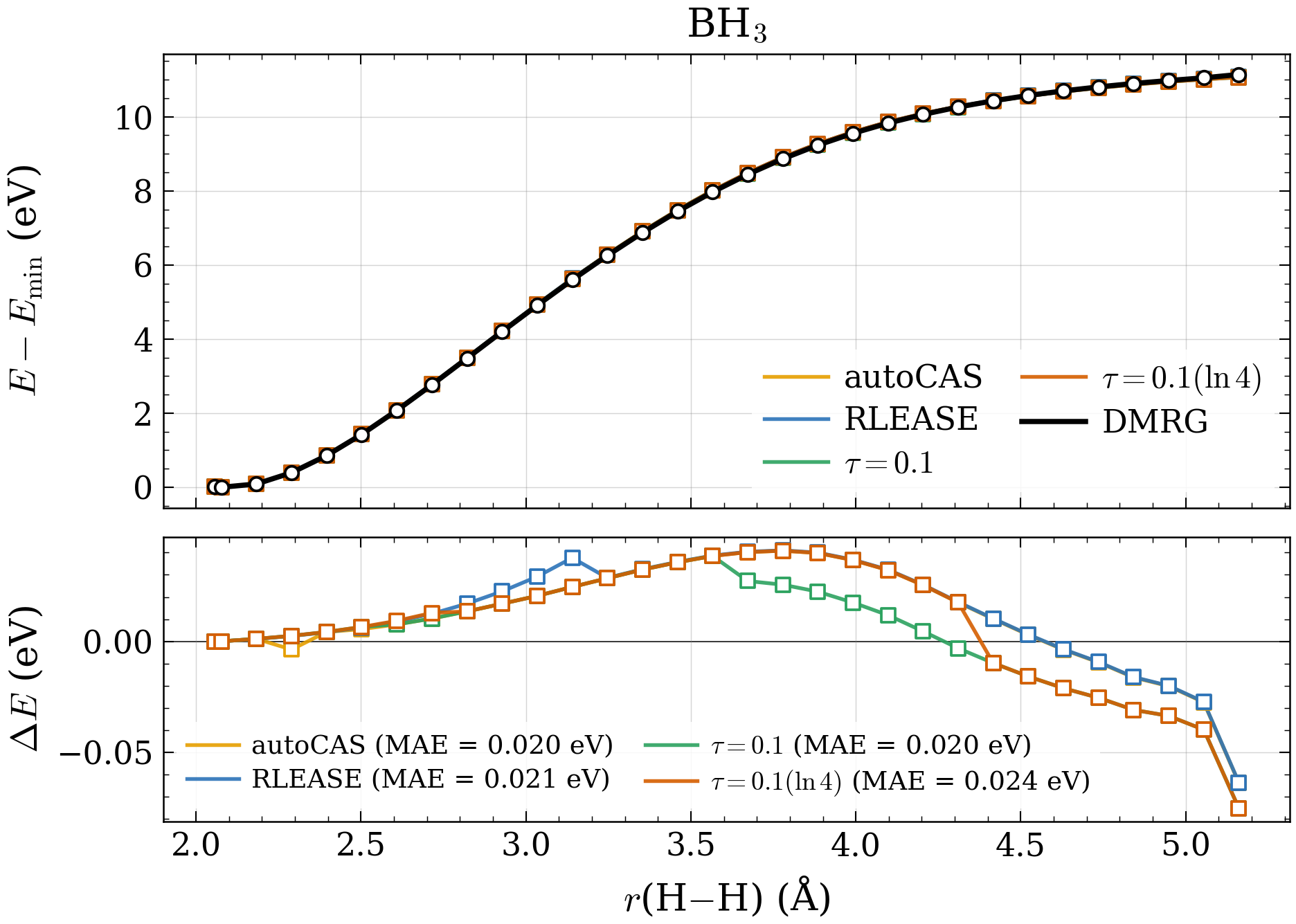}\\[4pt]
  \includegraphics[width=0.48\textwidth]{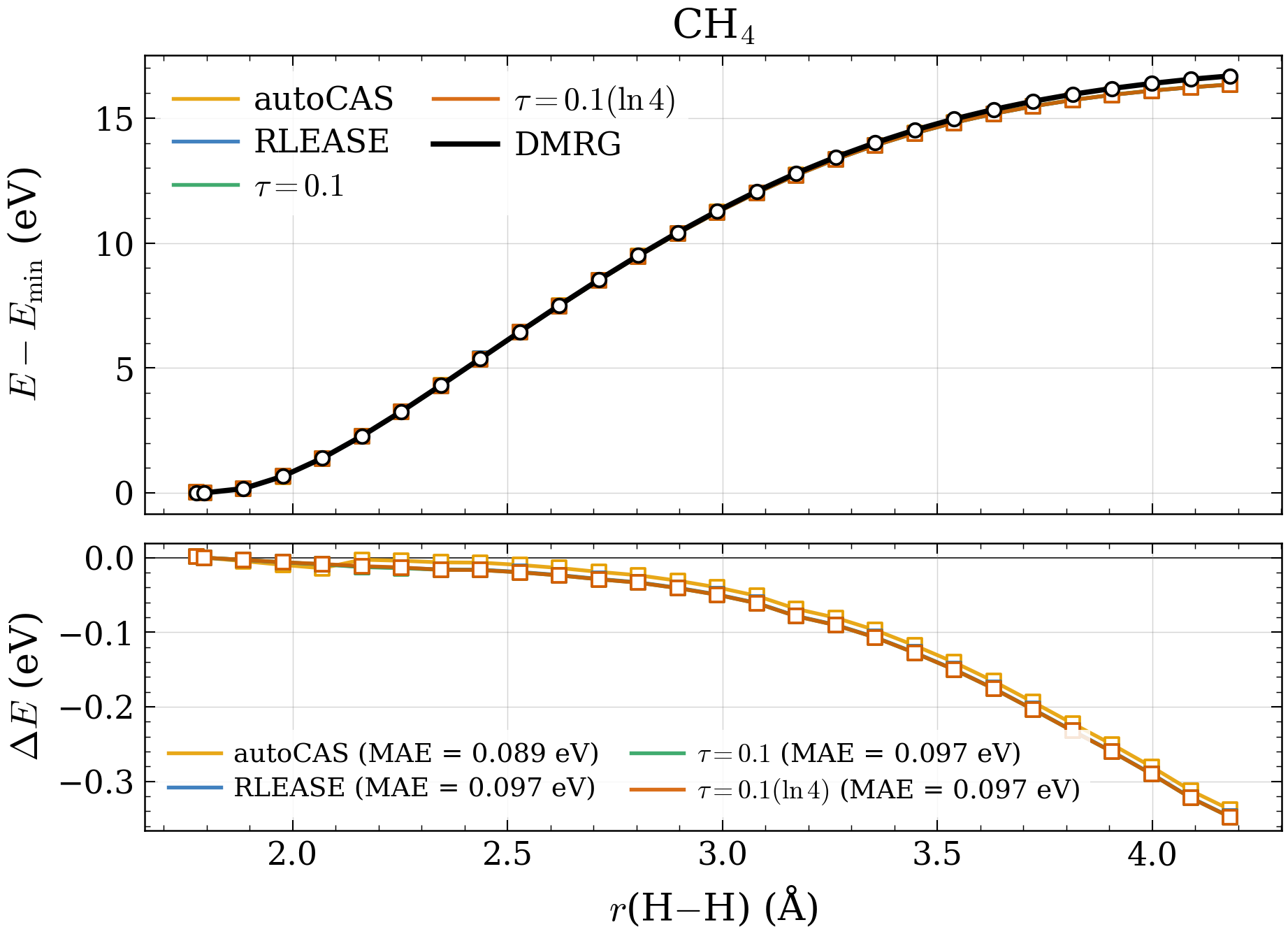}\hfill
  \includegraphics[width=0.48\textwidth]{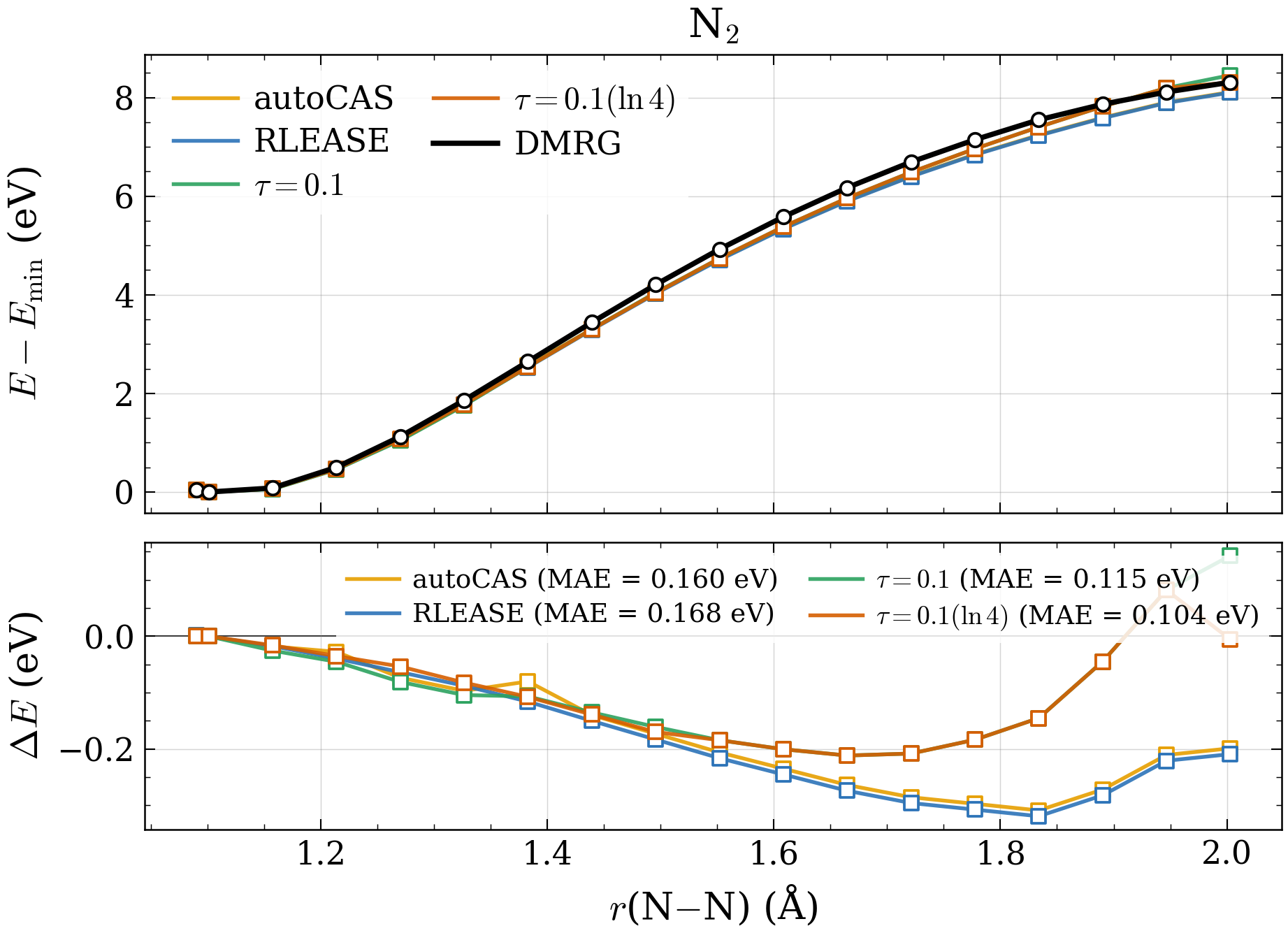}\\[4pt]
  \includegraphics[width=0.48\textwidth]{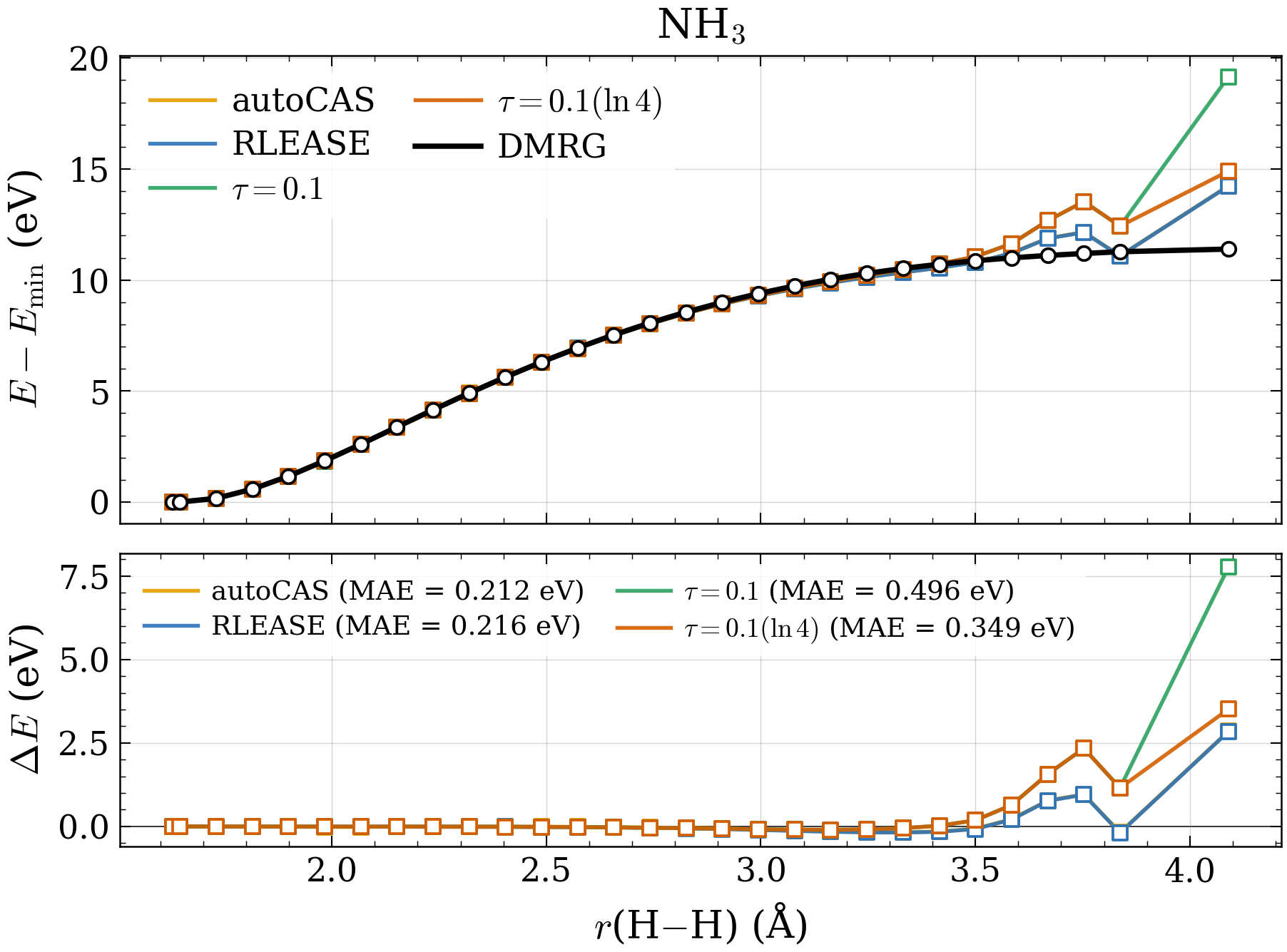}\hfill
  \includegraphics[width=0.48\textwidth]{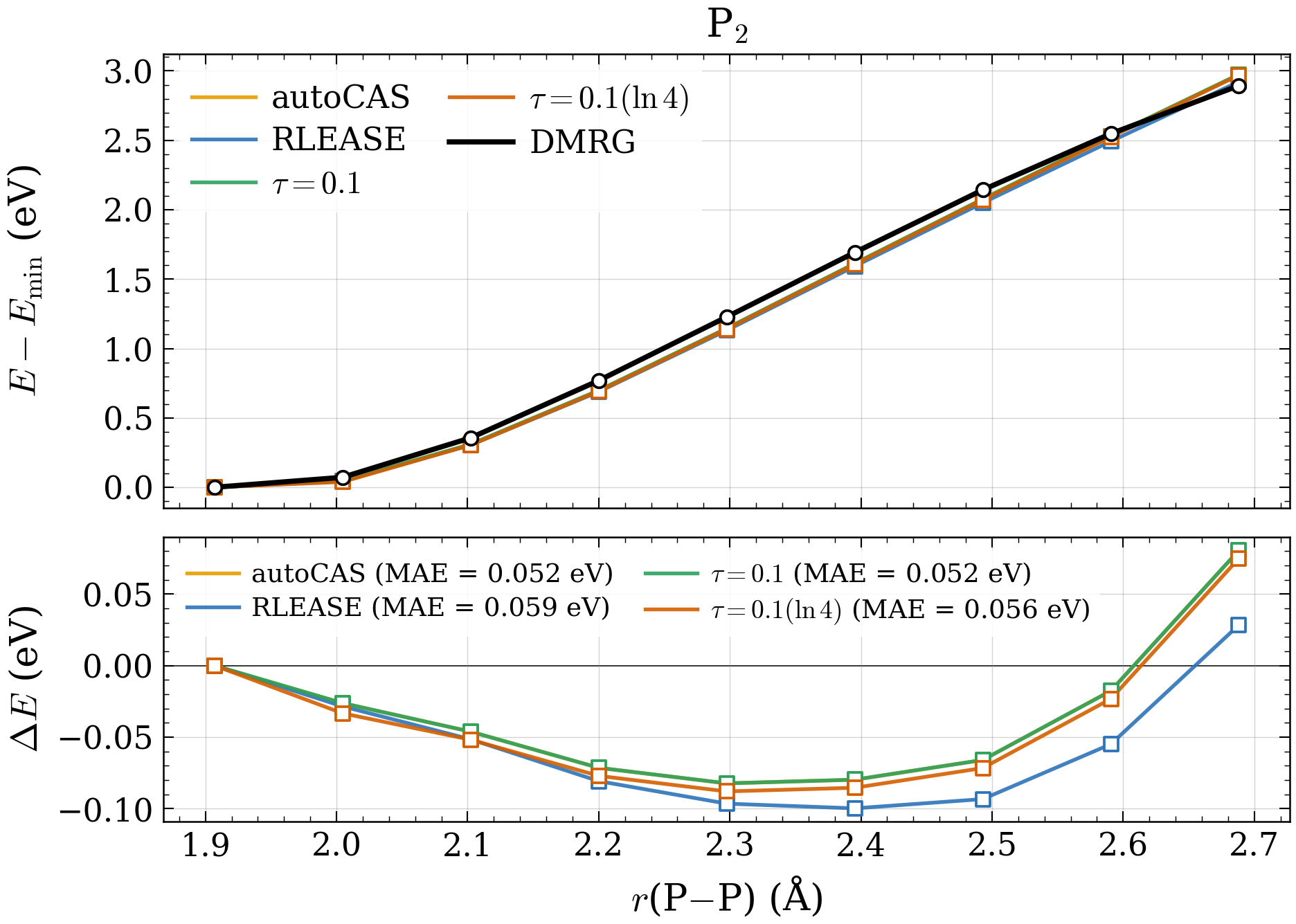}
  \caption{%
    Relative PES binding curves for ASF-CCSD(T) using active spaces
    selected by \rlease{}, autoCAS, $\tparam=0.1$, and
    $\tparam=0.1\ln 4$, compared to \dmrg{} ($D=1500$, cc-pVDZ)
    reference energies (black).
    The inclusion of perturbative triples substantially reduces
    errors near equilibrium relative to ASF-CCSD
    (see \cref{tab:relative-energy-mae}), while the active-space
    correction generally preserves accuracy in the stretched-bond regime
    where bare CCSD(T) amplitudes diverge.
  }
  \label{fig:pes_asfccsdt}
\end{figure}

An active space is selected to reproduce a target
property, so the quality of the selection should be judged by
how accurately that property is recovered.
In this work, the target property is the relative energy along a
potential energy surface.
Binding curves for the six molecules in the method-comparison subset
(BeH$_2$, BH$_3$, CH$_4$, N$_2$, NH$_3$, and P$_2$) and all three
downstream methods are shown in
\cref{fig:pes_nevpt2,fig:pes_asfccsd,fig:pes_asfccsdt}.
\Cref{tab:relative-energy-mae} reports the corresponding mean absolute
errors (MAEs) of relative PES energies with respect to \dmrg{}
($D=1500$, cc-pVDZ) for \rlease{}, autoCAS, and the two fixed
entropy-threshold baselines across all three downstream methods.

\begin{table}[ht]
\centering
\caption{%
  Mean absolute errors (MAE, eV) for relative PES energies computed
  against \dmrg{} ($D=1500$, cc-pVDZ).
  Results are shown for \rlease{}, autoCAS, and fixed entropy
  thresholds $\tparam=0.1$ and $\tparam=0.1\ln 4$.
  The overall row averages over all molecules in the subset.
}
\label{tab:relative-energy-mae}
\begin{tabular}{llrrrr}
\toprule
Method & Molecule & \rlease{} & autoCAS & $\tau=0.1$ & $\tau=0.1\ln 4$ \\
\midrule
ASF-CCSD      & Overall & \textbf{0.333} & 0.357 & 0.348 & 0.345 \\
              & BeH$_2$ & 0.055 & 0.124 & 0.067 & 0.083 \\
              & BH$_3$  & 0.202 & 0.199 & 0.179 & 0.188 \\
              & CH$_4$  & 0.224 & 0.254 & 0.217 & 0.218 \\
              & N$_2$   & 0.616 & 0.643 & 0.644 & 0.644 \\
              & NH$_3$  & 0.662 & 0.679 & 0.735 & 0.714 \\
              & P$_2$   & 0.237 & 0.243 & 0.243 & 0.225 \\
\midrule
ASF-CCSD(T)   & Overall & 0.103 & \textbf{0.101} & 0.141 & 0.116 \\
              & BeH$_2$ & 0.057 & 0.071 & 0.063 & 0.063 \\
              & BH$_3$  & 0.021 & 0.020 & 0.020 & 0.024 \\
              & CH$_4$  & 0.097 & 0.089 & 0.097 & 0.097 \\
              & N$_2$   & 0.168 & 0.160 & 0.115 & 0.104 \\
              & NH$_3$  & 0.216 & 0.212 & 0.496 & 0.349 \\
              & P$_2$   & 0.059 & 0.052 & 0.052 & 0.056 \\
\midrule
\nevpt{}      & Overall & \textbf{0.120} & 0.221 & 0.282 & 0.178 \\
              & BeH$_2$ & 0.180 & 0.173 & 0.270 & 0.212 \\
              & BH$_3$  & 0.125 & 0.097 & 0.375 & 0.326 \\
              & CH$_4$  & 0.243 & 0.482 & 0.499 & 0.310 \\
              & N$_2$   & 0.109 & 0.107 & 0.131 & 0.091 \\
              & NH$_3$  & 0.026 & 0.367 & 0.312 & 0.081 \\
              & P$_2$   & 0.039 & 0.102 & 0.102 & 0.049 \\
\bottomrule
\end{tabular}
\end{table}

Because the RL reward is derived from \nevpt{} energies, one expects
\rlease{} to be most strongly calibrated for the \nevpt{} pathway, and
this is borne out by \cref{tab:relative-energy-mae}: \rlease{} achieves
the lowest overall \nevpt{} MAE (0.120~eV), nearly halving the autoCAS
error (0.221~eV) and substantially outperforming both fixed thresholds.
The improvements are largest for CH$_4$ and NH$_3$, where autoCAS
selects much larger active spaces (\cref{fig:size_diff}) that do
not translate into better NEVPT2 energies.
Remarkably, the advantage transfers to downstream methods not seen
during training.
For ASF-CCSD(T), \rlease{} closely matches the autoCAS MAE (0.103 vs.\ 0.101~eV),
and both substantially outperform the fixed thresholds;
for ASF-CCSD the \rlease{} MAE is likewise the lowest overall.
The central result is that \rlease{} achieves the best overall
\nevpt{} MAE, closely matches autoCAS for ASF-CCSD(T), and does so
\emph{without} requiring a pilot \dmrg{} calculation in the target
basis to obtain orbital entropies; autoCAS and the fixed-threshold
baselines all depend on such a calculation.
\rlease{} requires only a Hartree--Fock calculation plus
millisecond-scale neural-network inference at deployment.
The method is therefore not merely reproducing the active spaces of
existing selectors; it selects compact active spaces that yield
competitive or improved downstream energies at substantially lower
selection cost.

Because the RL action space is a single scalar threshold, one might
ask whether simpler optimization strategies (grid search over
$\tparam$, Bayesian optimization) could achieve comparable results.
Two factors favor the PPO formulation.
First, each reward evaluation requires a full CASCI + NEVPT2
calculation, making exhaustive search over $\tparam$ expensive;
PPO amortizes this cost by learning from stochastic samples across
geometries and epochs.
Second, the optimal threshold depends on the molecular state (the
predicted $\hat{s}_1$ profile), which varies across molecules and
geometries; a single fixed threshold found by grid search may not
generalize, whereas the PPO-trained policy learns a distribution
that accounts for this variability.
A systematic comparison against alternative optimizers is a useful
direction for future work.

\FloatBarrier
\subsection{RLEASE binding curves for representative molecules}
\label{sec:results:rlease_only}

\begin{figure}[ht]
  \centering
  \includegraphics[width=0.48\textwidth]{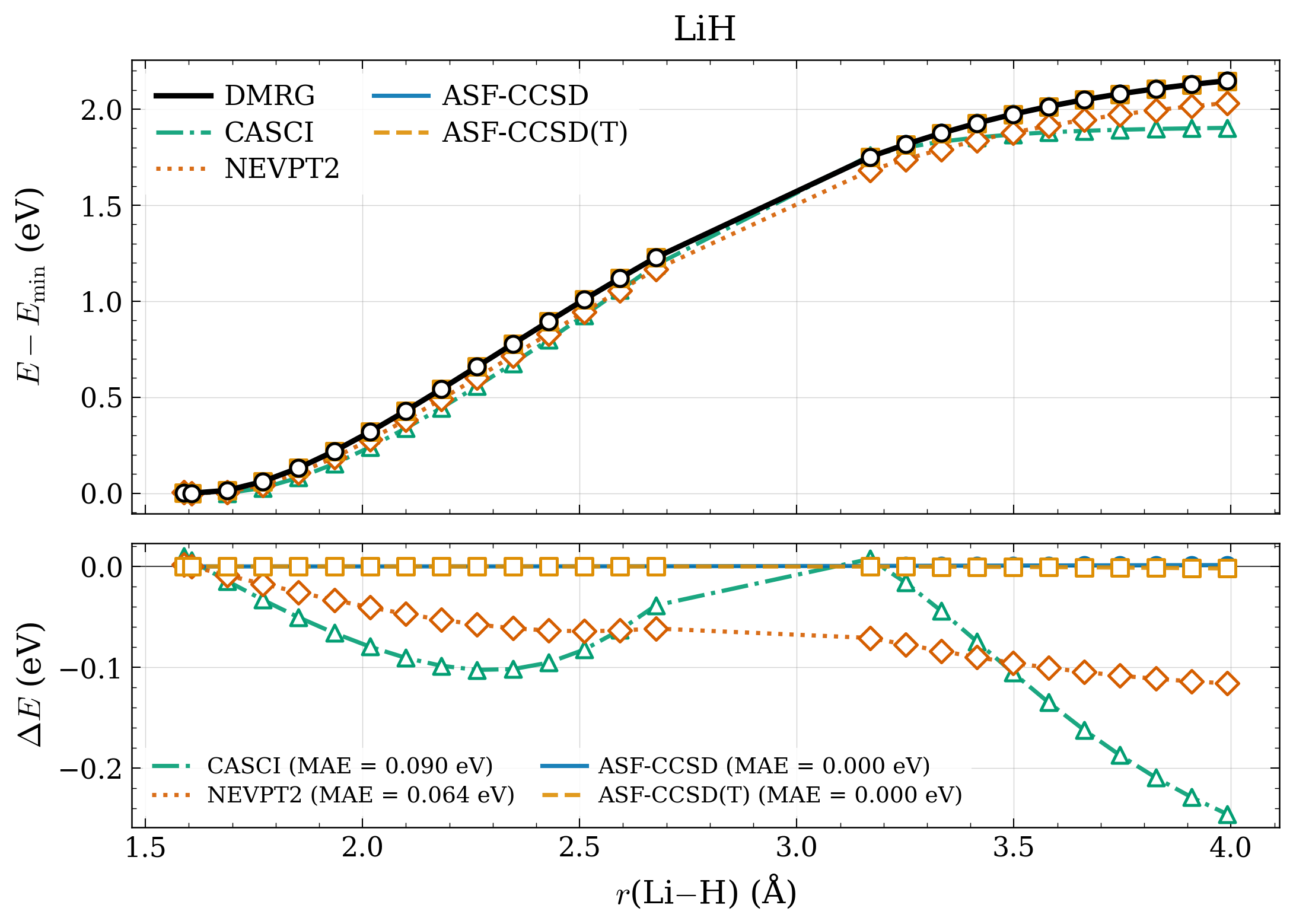}\hfill
  \includegraphics[width=0.48\textwidth]{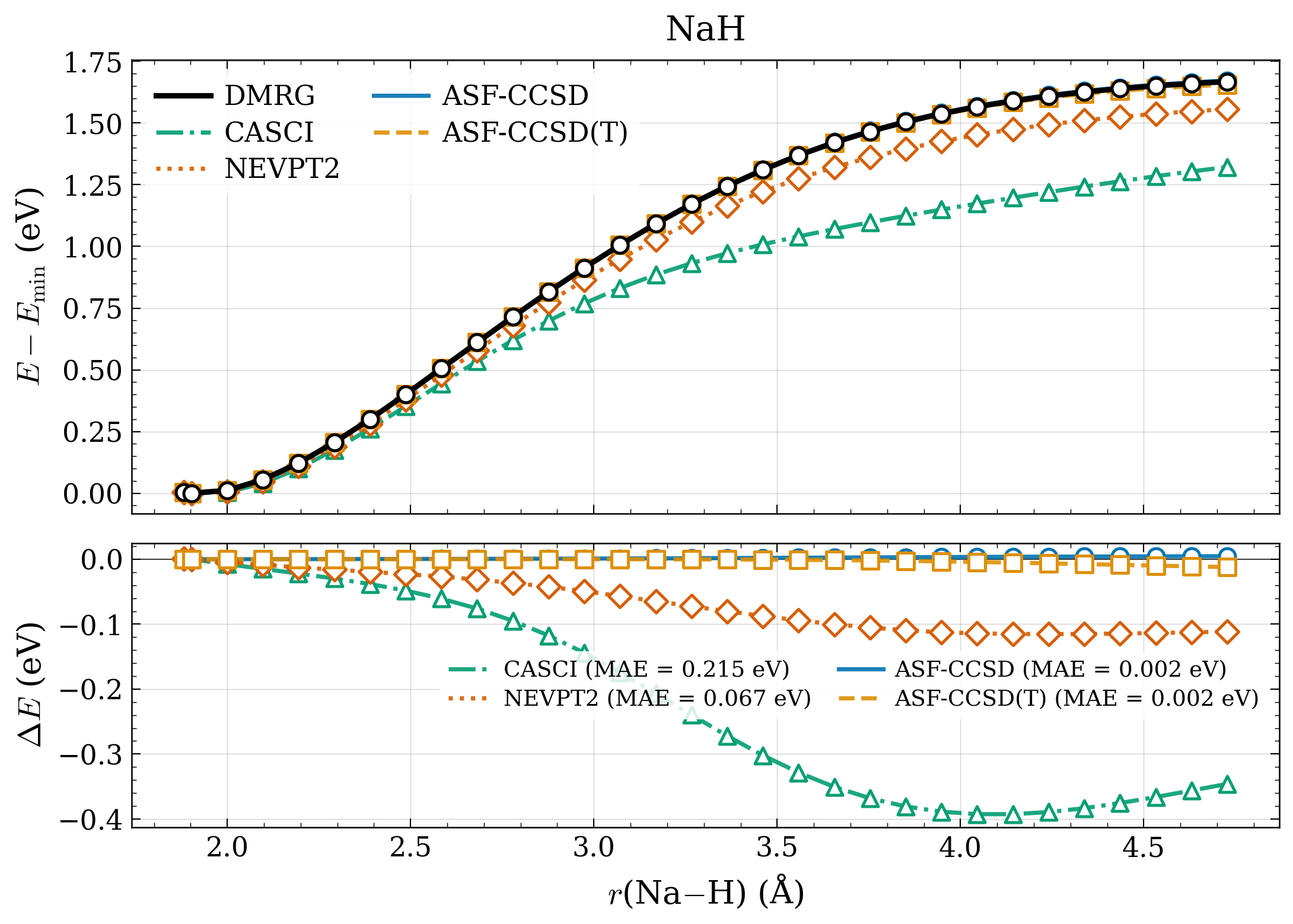}\\[4pt]
  \includegraphics[width=0.48\textwidth]{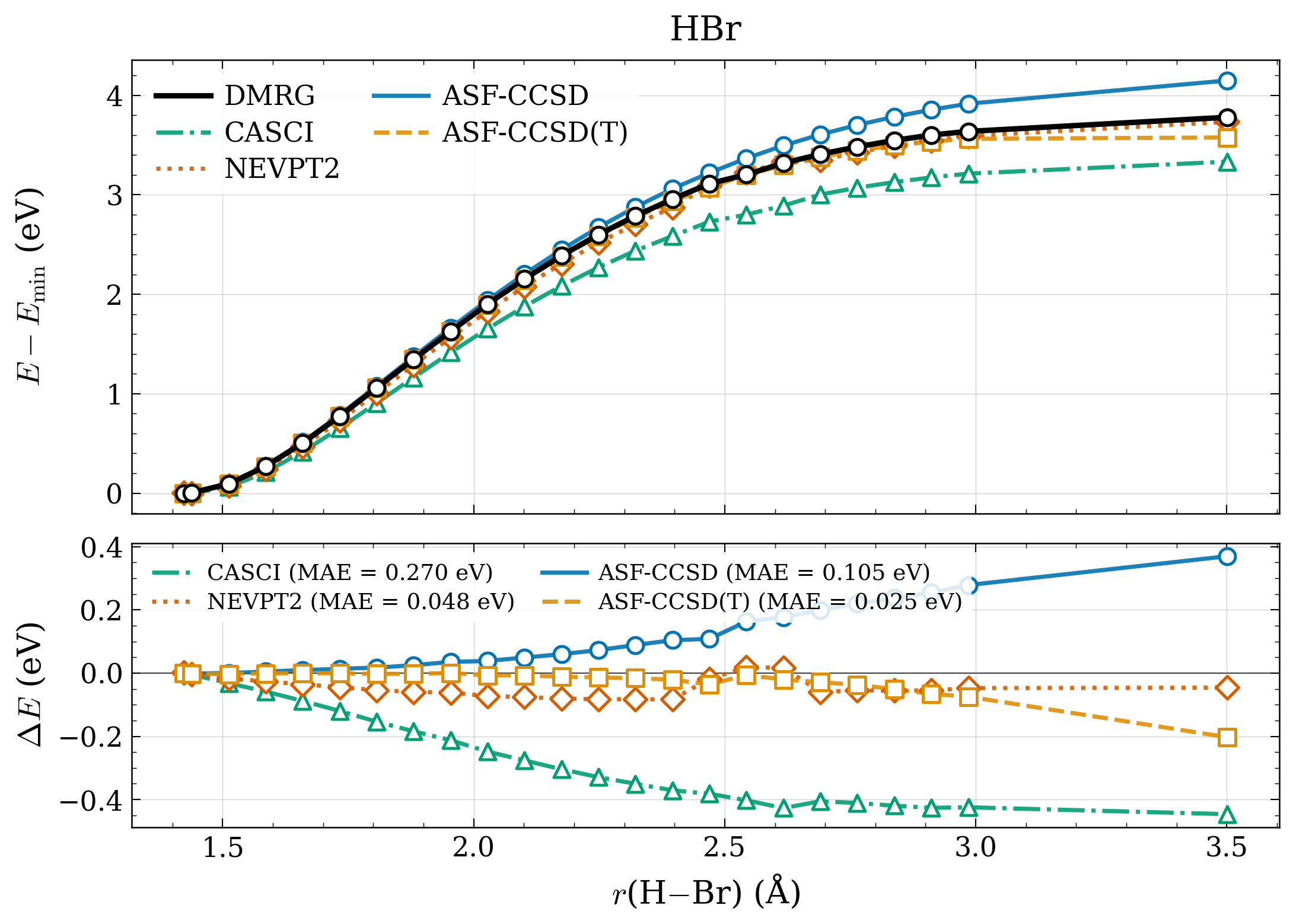}\hfill
  \includegraphics[width=0.48\textwidth]{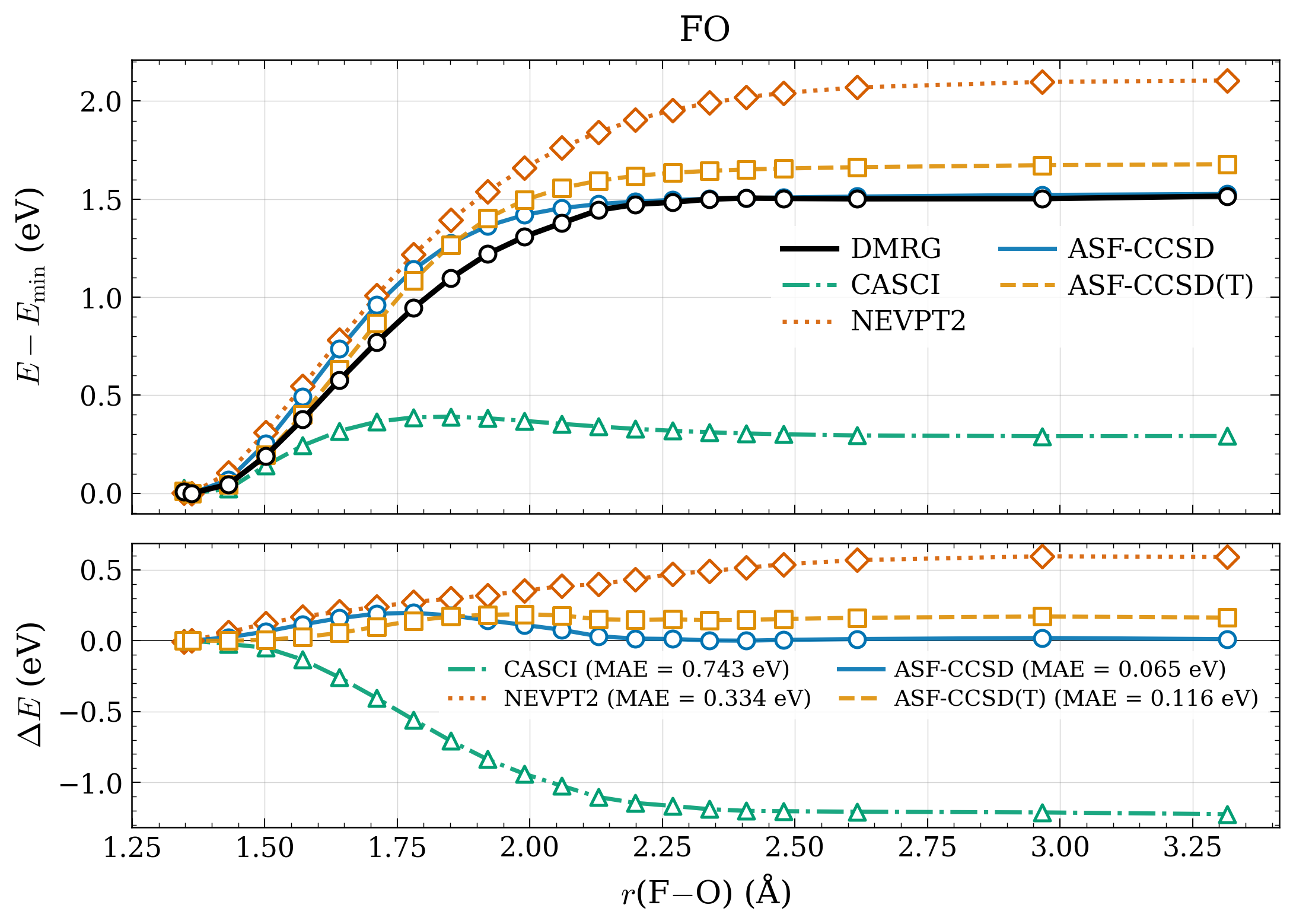}\\[4pt]
  \includegraphics[width=0.48\textwidth]{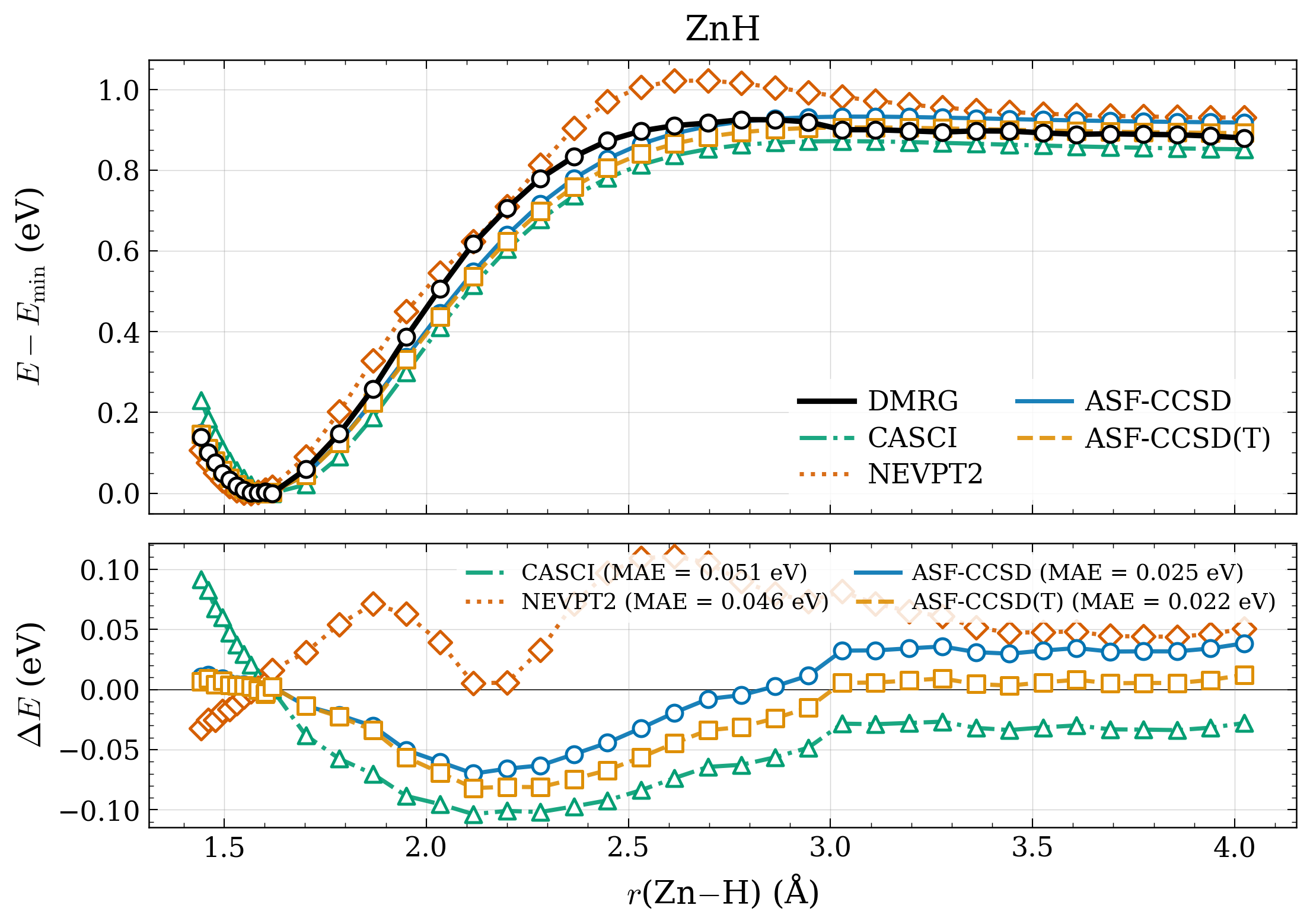}\hfill
  \includegraphics[width=0.48\textwidth]{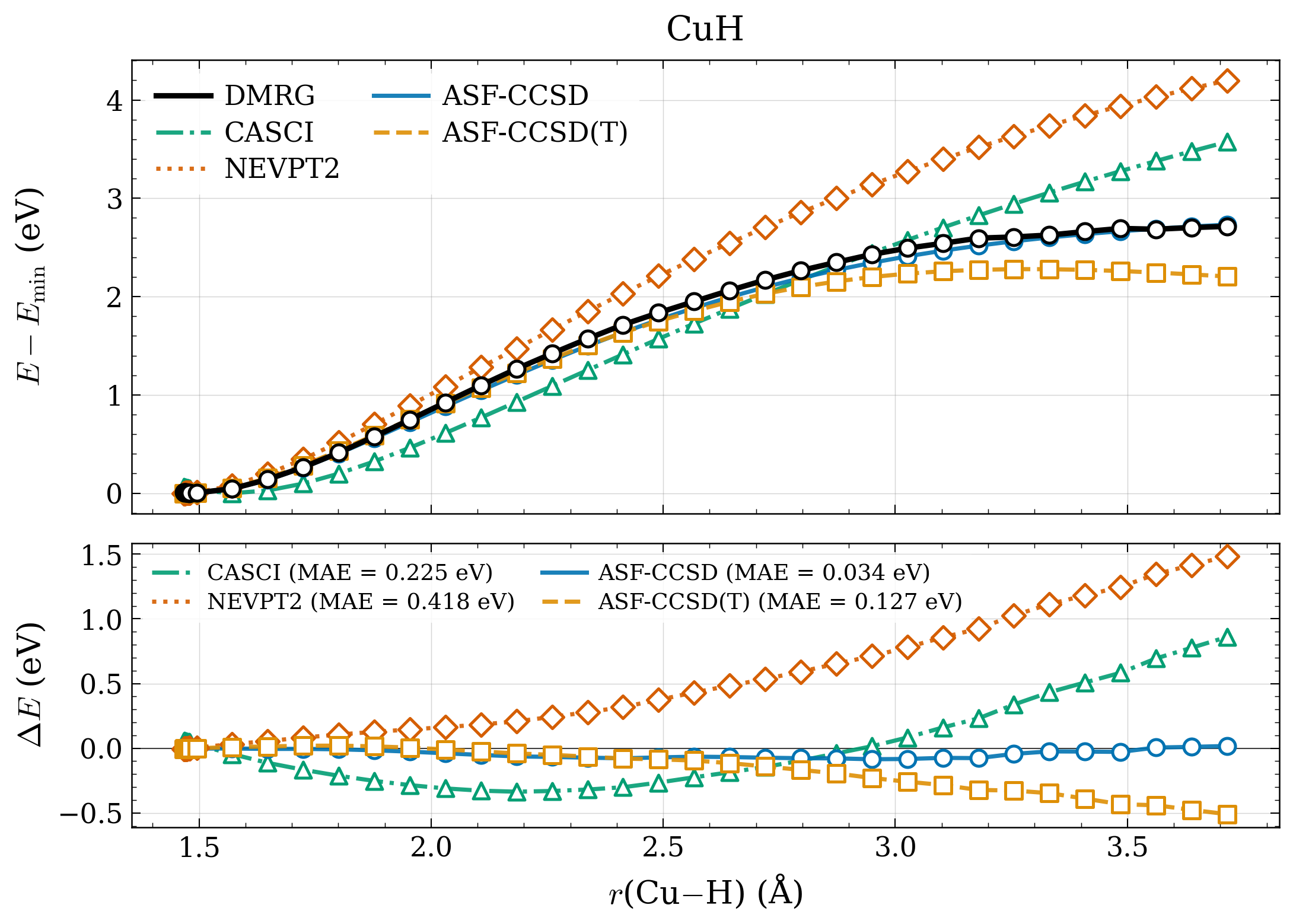}
  \caption{%
    Relative PES binding curves for \rlease{}-selected active spaces
    compared to \dmrg{} ($D=1500$, cc-pVDZ) reference energies (black),
    for four downstream methods: CASCI, \nevpt{}, ASF-CCSD, and ASF-CCSD(T).
    Top row: main-group hydrides LiH and NaH.
    Middle row: HBr and the open-shell FO radical.
    Bottom row: 3$d$ transition-metal hydrides ZnH and CuH.
    These molecules were not included in the method-comparison subset
    of \cref{fig:pes_nevpt2,fig:pes_asfccsd,fig:pes_asfccsdt} and
    demonstrate the transferability of the \rlease{} active-space
    selection across diverse chemical environments.
  }
  \label{fig:pes_rlease_only}
\end{figure}

\Cref{fig:pes_rlease_only} shows binding curves for six molecules
evaluated with the \rlease{}-selected active space across all three
downstream methods.
The top row covers the main-group hydrides LiH and NaH, which span different degrees of bond polarity. The middle row includes an additional main-group hydride HBr and the open-shell FO radical, the latter exhibiting pronounced open-shell multireference character. The bottom row shows the 3$d$ transition-metal hydrides ZnH and CuH, which require active spaces capable of capturing $d$-shell participation.
None of these molecules appear in the training set, and all use the
same learned threshold without retraining.
The transition-metal hydrides (ZnH, CuH) are a particularly stringent
test: the training set contains no transition-metal species, yet
\rlease{} selects active spaces that include orbitals of
predominantly $3d$ and $4s$ character.
Because the continuous orbital descriptors (energies, integrals,
spatial extent) for these orbitals still fall within ranges that
the network can meaningfully extrapolate from, the learned threshold
correctly identifies them as correlated despite the absence of any
transition-metal examples during training.
Moreover, because the selection depends only on whether predicted
$\hat{s}_1$ values exceed the threshold rather than on their precise
magnitudes, moderate prediction errors for out-of-distribution
orbitals do not necessarily degrade the final active-space choice.

\FloatBarrier
\subsection{Case study: \texorpdfstring{$p$}{p}-benzyne}
\label{sec:results:pbenzyne}

\begin{figure}[ht]
  \centering
  \includegraphics[width=\columnwidth]{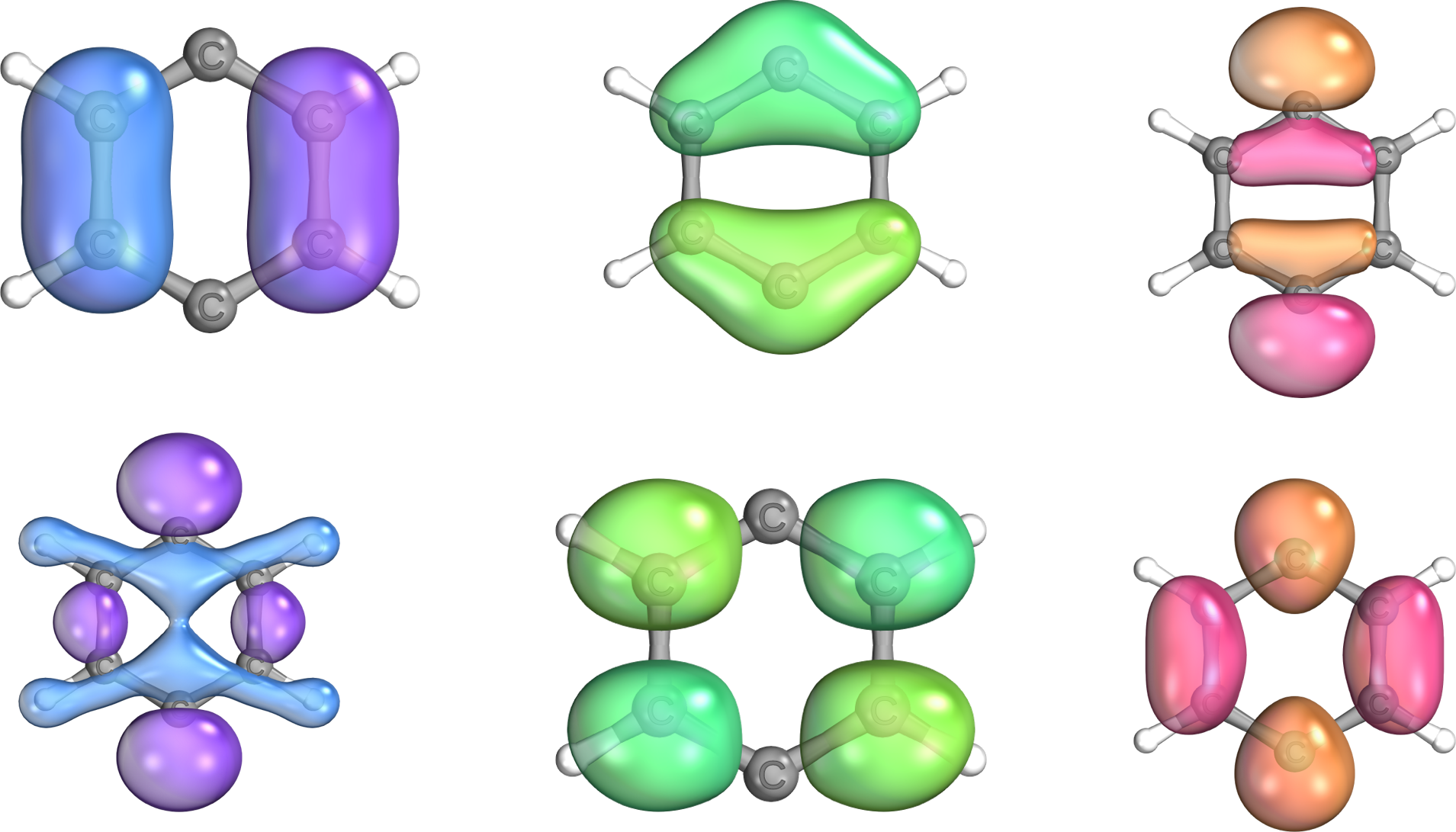}
  \caption{%
    The six active orbitals selected by \rlease{} for
    $p$-benzyne (1,4-didehydrobenzene) in the 6-31G* basis,
    yielding a CAS(6$e$,6$o$) active space.
  }
  \label{fig:pbenzyne}
\end{figure}

$p$-Benzyne (1,4-didehydrobenzene) is a prototypical
$\sigma{,}\sigma$ diradical that poses a well-known challenge to
single-reference methods because of the near-degeneracy of its
symmetric (S) and antisymmetric (A) radical
orbitals~\cite{ClarkDavidson2003}.
Clark and Davidson~\cite{ClarkDavidson2003} established, using
CASSCF in the 6-31G* basis, that a CAS(8$e$,8$o$) active
space, the six $\pi/\pi^*$ benzene-ring orbitals augmented by the
two $\sigma$ radical orbitals is the standard reference for this
system.

When deployed on $p$-benzyne without any molecule-specific
tuning, \rlease{} selects a CAS(6$e$,6$o$) active space (\cref{fig:pbenzyne}).
Inspection of the selected orbitals reveals a mixture of
$\pi/\pi^*$ ring orbitals and in-plane orbitals with $\sigma$
character localized on the 1,4 radical centers.
The full (8$e$,8$o$) reference space of Clark and
Davidson~\cite{ClarkDavidson2003} contains six $\pi/\pi^*$ and
two $\sigma$ radical orbitals, and their CASSCF natural orbitals exhibit strongly fractional occupations for the four $\pi/\pi^*$ and two $\sigma$ orbitals selected by \rlease{}, whereas the final two $\pi/\pi^*$ (not selected in our study) have near-integer occupation.
\rlease{} therefore captures only the essential $\pi$ static
correlation and the $\sigma{,}\sigma$ diradical character, providing
a compact yet physically meaningful active space.

This result is noteworthy because $p$-benzyne was absent from the
training set, which contains no aromatic or diradical species.
The fact that \rlease{} identifies orbitals with both $\pi$ and
$\sigma$-radical character as strongly entangled, using only HF
descriptors and the learned threshold, demonstrates that the model
has captured a transferable notion of orbital importance rather than
memorizing molecule-specific patterns.

A notable feature of these results is that the model was trained
entirely in the STO-3G basis yet deployed in 6-31G*.
This basis-set transfer succeeds because the orbital descriptors
used by \rlease{} (orbital energies, two-electron integrals, dipole
moments, and atomic-orbital composition) encode the \emph{qualitative
character} of each orbital (bonding vs.\ antibonding, $s/p/d$
angular-momentum content, spatial extent) rather than
basis-set-specific numerical values.
While the absolute magnitudes of these descriptors shift between
basis sets, the relative ordering and the distinction between
strongly and weakly correlated orbitals are largely preserved,
enabling the learned threshold to transfer without retraining.
\clearpage
\FloatBarrier
\section{Conclusions and Outlook}
\label{sec:conclusions}

We have introduced \rlease{}, a framework that reframes active-space
selection as a learned, energy-driven optimization problem.
By combining a neural-network predictor of per-orbital importance scores
with a PPO-optimized threshold, \rlease{} directly maximizes downstream
energy accuracy, rather than treating orbital selection as a
disconnected preprocessing step.

The main findings of this work are:
\begin{enumerate}
  \item \textbf{Energy-aware selection.}
    The RL-optimized threshold produces active-space selections that
    achieve overall MAE of 0.120~eV (\nevpt{}) and 0.103~eV
    (ASF-CCSD(T)) for relative PES energies against \dmrg{}
    ($D=1500$, cc-pVDZ), achieving the lowest overall MAE for
    \nevpt{} and ASF-CCSD and closely matching autoCAS for ASF-CCSD(T),
    while avoiding a target-basis pilot \dmrg{} at inference time.
  \item \textbf{Generalization from minimal training.}
    Trained on only three molecules (Na$_2$, ClF, SiO$_2$) and their
    PES geometries in the STO-3G basis, the model transfers without
    retraining to all test species shown in this work, including
    molecule types and a basis set (cc-pVDZ) not seen during training.
    Full four-method comparisons are reported for six molecules;
    RLEASE-only binding curves for six additional species
    (including 3$d$ transition-metal hydrides) provide further
    evidence of transferability, though broader benchmark coverage
    remains a direction for future work.
  \item \textbf{Method-agnostic deployment.}
    A single learned active space serves three complementary
    downstream methods (\nevpt{}, ASF-CCSD, and ASF-CCSD(T)) without
    requiring coupled-cluster calculations during training.
  \item \textbf{Instantaneous deployment.}
    Descriptor extraction and neural-network inference add $<1$~s to
    any new geometry, making \rlease{} compatible with high-throughput
    screening and molecular dynamics workflows.
\end{enumerate}

Several directions for future work emerge naturally.
First, replacing HF orbitals with CASSCF-optimized orbitals within the
RL loop would improve both the quality of the reference state and the
stability of orbital orderings, particularly for early transition metals
at stretched geometries.
Second, extension to larger basis sets (cc-pVTZ, aug-cc-pVDZ) will test
basis-set transferability of the learned descriptors.
Third, application to reaction paths and transition-state geometries
would demonstrate utility for catalysis workflows, where active spaces
change qualitatively along the reaction coordinate.
Finally, coupling \rlease{} with ML interatomic potentials
(e.g.\ MACE~\cite{Batatia2022mace} and
NequIP~\cite{Batzner2022nequip}) could enable on-the-fly
multireference molecular dynamics with automatically selected
active spaces at each time step.

\FloatBarrier
\begin{acknowledgments}
\textbf{Author contributions.}
E.O. initiated the RLEASE research direction, performed the DMRG calculations and analysis, and implemented the agent active-space selection workflow. E.O., A.M., A.J.J., A.Ka., and D.R. contributed to the conception and development of the RLEASE methodology. A.M., A.J.J., and D.R. participated in the electronic-structure calculations. E.O. and A.Ku. generated the training and test data structures. E.O., K.A.P., and A.M. generated the orbital descriptors. V.A.N. and R.H.L. contributed to the electronic-structure component of the work. A.Ka. and D.R. contributed to project coordination and supervision. E.O., A.M., A.J.J., and D.R. wrote the initial manuscript draft; all authors participated in manuscript revision.

\textbf{Acknowledgments.}
The authors acknowledge the broader PsiQuantum application team for useful
discussions on multi-reference electronic structure and active-space
selection.
\end{acknowledgments}
\clearpage
\section*{References}
\bibliographystyle{apsrev4-2}
\bibliography{references}

\end{document}